%% file: green_valley.tex
\documentclass[a4paper,fleqn,usenatbib]{mnras}

\usepackage{graphicx}
\usepackage{float}
\usepackage{acronym}
\usepackage{aas_macros}
\usepackage{amssymb}
\usepackage{amsfonts}
\usepackage{xcolor}
\usepackage{hyperref}

\newcommand{\eagle}[0]{{\sc eagle}}

\newcommand{\us}{{\it u}$^\star$}
\newcommand{\rs}{{\it r}$^\star$}
\def\aap{A\&A}
\def\apj{ApJ}
\def\mnras{MNRAS}

\title[Colour evolution in EAGLE]{It's not easy being green: The evolution of galaxy colour in the EAGLE simulation}

\author[J.W. Trayford, et al.]{James W. Trayford$^{1}$\thanks{E-mail: j.w.trayford@durham.ac.uk (JWT)}, Tom Theuns$^1$, 
Richard G. Bower$^1$, Robert A. Crain$^2$,
\newauthor
 Claudia del P. Lagos$^{3,4}$, Matthieu Schaller$^1$, Joop Schaye$^5$\\
$^{1}$Institute for Computational Cosmology, Durham University, South Road, Durham, DH1 3LE.\\
$^2$Astrophysics Research Institute, Liverpool John Moores University, 146 Brownlow Hill, Liverpool, L3 5RF.\\
$^3$International Centre for Radio Astronomy Research, 7 Fairway, Crawley, 6009, Perth, WA, Australia.\\
$^4$Australian Research Council Centre of Excellence for All-sky Astrophysics (CAASTRO), 44 Rosehill Street Redfern,\\
 NSW 2016, Australia.\\
$^5$Leiden Observatory, Leiden University, P.O. Box 9513, 2300 RA Leiden, the Netherlands.
}

\begin{document}

\date{Accepted . Received ; in original form }

\pagerange{\pageref{firstpage}--\pageref{lastpage}} \pubyear{2015}

\maketitle
\label{firstpage}

\begin{abstract}
 We examine the evolution of intrinsic $u$-$r$ colours of galaxies in the \eagle{} cosmological hydrodynamical simulations, which has been shown to reproduce the observed redshift $z=0.1$ colour-magnitude distribution well, with a focus on $z < 2$. The median $u$-$r$ of star-forming (\lq blue cloud\rq) galaxies reddens by 1~magnitude from $z=2$ to $0$ at fixed stellar mass, as their specific star formation rates decrease with time. A red sequence starts to build-up around $z=1$, due to the quenching of low-mass satellite galaxies at the faint end, and due to the quenching of more massive central galaxies by their active galactic nuclei (AGN) at the bright end. This leaves a dearth of intermediate-mass red sequence galaxies at $z=1$, which is mostly filled in by $z=0$. We quantify the time-scales of colour transition finding that most galaxies spend less than 2~Gyr in the \lq green valley\rq. We find the timescale of transition to be independent of quenching mechanism, i.e. whether a galaxy is a satellite or hosting an AGN. On examining the trajectories of galaxies in a colour-stellar mass diagram, we identify three characteristic tracks that galaxies follow (quiescently star-forming, quenching and rejuvenating galaxies) and quantify the fraction of galaxies that follow each track.
\end{abstract}


\input{acronyms}

\input{Introduction}
\input{Simulation}

\input{Evolution}

\input{Quenching}
\input{Summary}
\section*{Acknowledgements}

The authors would like to thank Adam Muzzin for insightful discussion of this work and Michelle Furlong for providing valuable comments on an early draft of the manuscript. We would also like to thank the anonymous referee, who suggested revisions from which the manuscript greatly benefited. This work was supported by the Science and Technology Facilities Council [grant number ST/F001166/1], by the Interuniversity Attraction Poles Programme initiated by the Belgian Science Policy Office ([AP P7/08 CHARM]) by ERC grant agreement 278594 - GasAroundGalaxies, and used the DiRAC Data Centric system at Durham University, operated by the Institute for Computational Cosmology on behalf of the STFC DiRAC HPC Facility (www.dirac.ac.uk). This equipment was funded by BIS National E-Infrastructure capital grant ST/K00042X/1, STFC capital grant ST/H008519/1, and STFC DiRAC is part of the National E-Infrastructure. RAC is a Royal Society University Research Fellow. The data used in the work is available through collaboration with the authors. CL is funded by a Discovery Early Career Researcher Award (DE150100618). We acknowledge the Virgo Consortium for making their simulation data available. The \eagle{} simulations were performed using the DiRAC-2 facility at Durham, managed by the ICC, and the PRACE facility Curie based in France at TGCC, CEA, Bruy\`{e}res-le-Ch\^{a}tel.

\bibliographystyle{mnras}
\bibliography{references}

\end{document}

%% file: acronyms.tex
\acrodef{CMD}{colour magnitude diagram}
\acrodef{LF}{luminosity function}
\acrodef{GAMA}{Galaxy And Mass Assembly}
\acrodef{UKIDSS}{UK Infra-red Digital Sky Survey}
\acrodef{SDSS}{Sloan Digital Sky Survey}
\acrodef{RT}{Full 3-dimensional Radiative Transfer}
\acrodef{SED}{Spectral Energy Distribution}
\acrodef{T15}{\citet{Trayford15}}
\acrodef{K98}{\citet{Kennicutt98}}
\acrodef{S15}{\citet{Schaye15}}
\acrodef{C15}{\citet{Cheng15}}
\acrodef{D4000}{the 4000\AA{} break}


\acrodef{Ref-100}{Ref\_{}L0100N1504}
\acrodef{Recal-25}{Recal\_{}L0025N0754}
\acrodef{Ref-25}{Ref\_{}L0025N0376}
\acrodef{ML}{maximum-likelihood}
\acrodef{MCMC}{Markov-chain Monte Carlo}

%% file: Introduction.tex
\section{Introduction}
A scatter plot of observed galaxies in optical colour versus broad-band magnitude (or stellar mass) reveals two relatively well-defined distinct populations, a \lq red sequence\rq\ and a \lq blue cloud\rq, in a volume-limited sample. While a narrow red sequence was evident in early datasets \citep[e.g.][]{Sandage78, Larson80, Bower91}, this striking colour \lq bi-modality\rq\ was perhaps first revealed most clearly by \cite{Strateva01}, who exploited the step-change in sample size offered by the Sloan Digital Sky Survey ({\sc sdss}, \citealt{York00}), the statistics and general properties of these two sequences were subsequently characterised quantitatively by e.g. \cite{Baldry04}.
The {\it u-r} colour of a galaxy correlates strongly with its morphology (Hubble type); the {\sc galaxy zoo}\footnote{\url{http://www.galaxyzoo.org/}} citizen science project and the {\sc MegaMORPH} survey have put this correlation on a firm statistical footing \citep{Willet13, Haussler13}. Blue colours are typically due to light from massive hot, young stars. Broadly speaking, the more massive blue galaxies are star-forming discs and the fainter ones are irregulars. The red galaxies, in contrast, are elliptical or lenticular types, with the red colour reflecting an old stellar population. These galaxies are often referred to as \lq red-and-dead\rq\ to stress that their star formation has mostly ceased \citep[e.g.][]{Brammer09}.

At higher redshifts, $z\ge 1$, the blue cloud is clearly dominant and its colour becomes increasingly blue with increasing $z$. Selection effects bias against the detection of red galaxies at higher $z$, but it is clear that the bright end of the red sequence is already in place by $z\sim 1$, albeit with a small shift in colour \cite[e.g.][]{Wolf03, Bell04}

It is difficult to establish how these sequences arise, or how individual galaxies evolve in colour space, using observations alone. This is because both star formation and galaxy destruction by mergers change the number density of galaxies of given mass across time. \cite{Faber07} noted that the number density of blue galaxies remains approximately constant below redshift $z\sim 1$ whereas that of red galaxies increases markedly. This led them to propose a model in which one or more mechanisms operate that decrease the star formation rate of blue galaxies, with such \lq quenching\rq\ making galaxies redder until they join the red sequence. \cite{Bell12} showed that there is significant scatter in the properties of quenched galaxies. One correlation that stood out in their sample, is that quenched galaxies usually exhibit a prominent bulge, by association suggesting a supermassive black hole. 

A model in which an accreting supermassive black hole quenches star formation in its host galaxy appears very attractive. This is because black hole mass increases rapidly as a function of bulge mass \cite[e.g.][]{Haring04, McConnell13}, hence such a model might explain why most massive galaxies are red - as observed. Unfortunately the evidence that star formation in galaxies hosting X-ray bright AGN is indeed suppressed appears inconclusive. Several studies have found no correlation between star formation rate and X-ray luminosity for an X-ray selected sample of AGN \citep[e.g.][]{Rosario12, Harrison12, Stanley15}. However a close to linear correlation has been observed for galaxies selected in the infrared \citep[e.g.][]{Delvecchio15}. The fact that the luminosity of an AGN likely {\em varies} on a range of time-scales (from hours to Myrs) might explain the apparent disparity \citep{Hickox14, Volonteri15}. Powerful radio galaxies associated with the centres of groups and clusters do appear to disrupt the inflow of cold gas \citep{McNamara12}.

Another well-documented process that quenches star formation in a galaxy is restriction of its supply of gas by either ram-pressure stripping of disc gas \citep[e.g.]{Gunn72} or removal of halo gas \citep[e.g. strangulation][]{Larson80}, as the galaxy traverses a region of higher gas pressure associated with a group or cluster. The quenching of star-formation turns these satellites red \citep[e.g.][]{Knobel13}. Originally suggested by \citet{Gunn72}, the efficiency of these mechanisms have been investigated using simulations by many groups \cite[e.g.][]{Quilis00, Roediger07}, with more recently \citet{Bahe13} pointing out that galaxies may be stripped {\em before} they become satellites, by the gas in the outskirts of massive systems. \citet{McCarthy08} presented a theoretical framework that improves upon the simple analysis by \citet{Gunn72}, and describes their simulation results well.

Observational confirmation that environmental quenching indeed operates is evidenced by the fact that red galaxies preferentially reside in regions of high galaxy number density \citep{Dressler80}, or equivalently that red galaxies are more strongly clustered than blue galaxies, even at fixed mass \citep[e.g.][]{Zehavi05}, and that the clustering amplitude of red galaxies depends little on mass \citep[in contrast to that of blue galaxies, e.g.][]{Coil08}. This is compatible with a model where red galaxies reside close to, or even inside, more massive and hence strongly clustered halos that cause the quenching. Trends between the environment and gas content of galaxies provide further evidence, with galaxies residing in clusters seen to be deficient in both H{\sc I} and H$_2$ gas relative to the field \citep[e.g.][]{Cortese11, Boselli14}. Particularly convincing is the similarity of the trails of H{\sc I} gas seen to be emanating from gas rich galaxies in clusters \cite[e.g.][]{Chung07, Fumagalli14} and of the ram-pressure stripped gas behind simulated galaxies that fall onto a cluster \citep[e.g.][]{Roediger08}.

Even though observations suggest two empirical models of quenching ({\em i.e.} AGN and environmental), 
models of galaxy formation have struggled to reproduce simultaneously the detailed distribution of galaxies in the colour-magnitude diagram and the different clustering properties of red and blue galaxies. This is true of semi-analytical models, which use phenomenological prescriptions to describe the physical processes that lead to quenching \citep[e.g.][]{Font08, Lacey15}; for example, \citet{Henriques15} compare the Munich semi-analytical {\sc L-galaxies} model to {\sc sdss} data. Although in many aspects this model reproduces the observations better than its predecessors, limitations remain. For example, {\sc L-galaxies}' $u-r$ colours are considerably more bimodal than observed.

Hydrodynamical simulations can, in principle, model many physical processes self-consistently, but lack of numerical resolution and other limitations of the hydrodynamical integration may limit their realism.
Fortunately, relatively small changes to the basic hydrodynamics scheme \citep[e.g.][]{Price08, Hopkins13} seem to resolve most numerical issues, such that the dominant uncertainties in hydrodynamical simulations become associated to the implementation of unresolved \lq subgrid\rq\ processes rather than the details of the hydrodynamics scheme \citep{Scannapieco12, Schaller15}.

The huge dynamic range required to simulate a cosmologically representative volume with the required resolution to follow the hierarchical build-up of galaxies, presents a major challenge to numerical simulations. Until recently, such simulations did not reproduce the galaxy stellar mass function well, let alone the detailed colours/clustering of galaxies. A red/blue bimodality appears in the zoomed-simulations of \citet{Cen14} even though these do not include AGN. However, the $r-$band luminosity function of these simulation contains many more massive galaxies than observed. \citet{Gabor12} include the effects of AGN using a heuristic prescription of heating gas, where cooling is simply switched off in halos deemed massive enough to host AGN. They illustrate how this process builds-up a red sequence below redshift $z\sim 2$; initially lower-mass satellites and more massive quenched centrals appear in heated halos, with a characteristic dip in the abundance of red galaxies of stellar mass $M_\star\sim 10^{10}$~M$_\odot$ that is more prominent at higher $z$. While this simulation may provide valuable insight into the build up of the red sequence, the heuristic nature of the halo heating limits their practical applicability. For lower mass galaxies, \citet{Sales15} show that the {\sc illustris} simulation \citep{Vogelsberger14} broadly reproduces the colours of satellites, which they attribute to the relatively large gas fractions of satellites at infall.

The \eagle{} reference model was calibrated to the $z=0.1$ stellar mass function, black hole masses and sizes of galaxies and is currently the only hydrodynamical simulation that reproduces these observations. {\sc Eagle} also reproduces many independent galaxy observations, such as the content and ionisation state of gas \citep{Bahe16, Lagos15a}, mass profiles \citep{Schaller14} and evolution in stellar mass, star formation rate and size \citep{Furlong14, Furlong15}. The clustering of galaxies as a function of colour is investigated in a companion paper (Artale et al., {\em in prep.}). \citet{Trayford15} showed that \eagle{} reproduces the {\it g-r} $\--$ $M_r$ colour magnitude (and the {\it g-r} $\--$ $M_\star$) relation from the {\sc gama} spectroscopic survey \citep{Driver11} very well. Including a model for dust-reddening computed using the {\sc skirt} radiative transfer scheme \citep{Baes05, Camps15} improves the quantitative agreement further (Trayford et al. {\em in prep.}). With low-redshift ($z \sim 0.1$) galaxy colours in \eagle{} appearing to be realistic, studying how they have arisen given the physical feedback model of the simulation may provide new insight. The evolution of \eagle{} galaxy colours is also afforded credibility by the reasonable evolution of the \eagle{} galaxy population in terms of the stellar mass function \citep{Furlong14}. 

In section~\ref{sec:sim} we describe the \eagle{} simulations used in this study, particularly the aspects of star formation, metal enrichment and feedback that are most relevant for setting intrinsic galaxy colours. We investigate in section~\ref{sec:colev} the evolution of the galaxy population across the colour-mass diagram and correlate colour changes with galaxies becoming satellites or hosting an AGN. In section \ref{sec:quench} we expound these processes by analysing the behaviour of individual galaxies, using galaxy merger trees. Typical time-scales associated with colour transition are presented in section~\ref{sec:time}. We show that the colour evolution of most galaxies can be described well in terms of three generic tracks and quantify the fraction of galaxies that follow each path. Finally, our findings are summarised in section~\ref{sec:summary}. Throughout this work we refer to dust-free, rest-frame colours as \lq intrinsic\rq\ colours, and we take $Z_{\odot}=0.0127$ for the metallicity of the Sun \citep{Allende01}. Note that while the $Z_{\odot}$ value affects the normalisation of metallicities in solar units, colours are unaffected by the assumed $Z_{\odot}$ \citep[see][]{Trayford15}.

%% file: Simulation.tex
\section{The EAGLE simulations}
\label{sec:sim}

  The \eagle\ suite \citep{Schaye15, Crain15} includes simulations performed in a range of periodic volumes and at various numerical resolutions to enable convergence testing. The simulations were performed with the {\sc gadget-3} tree-SPH code \citep{Springel05}, but with changes to the SPH and time-stepping algorithm collectively referred to as {\sc anarchy} (see appendix of \cite{Schaye15} for details and \cite{Schaller15} for the relatively minor impact of these changes on the  properties of simulated galaxies). We use the $\Lambda$CDM cosmological parameters advocated by \cite{Planck}, and initial conditions generated at $z=127$ \citep{Jenkins13a} using second order Lagrangian perturbation theory. We concentrate here on analysing the largest reference model (Ref-100). This is a cubic cosmological volume of 100 comoving Mpc (cMpc) on a side, with an initial gas particle mass of $m_g = 1.81\times 10^6$~M$_\odot$. The simulation has a Plummer equivalent gravitational softening of $\epsilon_{\rm prop}=0.7 $ proper kpc (pkpc) at redshift $z=0$.

\subsection{Subgrid model and galaxy identification}
\label{sec:subgrid}
The \eagle{} reference model implements subgrid modules for physical processes that occur below the resolution limit, corresponding approximately to the Jeans length of the warm ISM. The free parameters that enter the modules for feedback were calibrated using the redshift $z=0.1$ galaxy stellar mass function, the $z=0.1$ stellar mass-size relation, and the $z=0$ stellar mass - black hole mass relation, see \cite{Crain15} for motivation and details. We briefly summarise these subgrid modules here, paying particular attention to those aspects most crucial for this paper.

\begin{itemize}
\item {\em Radiative cooling and photo-heating of gas} by the evolving optically-thin UV/X-ray background of \cite{Haardt01} is implemented element-by-element following \cite{Wiersma09a}.
\item {\em Star formation} is implemented as a pressure-law \citep{Schaye08} so that simulated galaxies reproduce the observed $z=0$ relation between gas and star formation surface density of \cite{Kennicutt98}. Each gas particle is assigned a star formation rate, $\dot{m}_\star$, and gas particles are converted to star particles stochastically. The star formation rate is zero for particles below the metallicity-dependent threshold of \cite{Schaye04}. We resample young stars using a probability proportional to the estimates of $\dot{m}_\star$ to reduce sampling noise when estimating the galaxy luminosities, as described in \citet{Trayford15}.
\item {\em Feedback from star formation} is implemented by heating gas particles neighbouring newly-formed star particles as described by \cite{DallaVecchia12}. In this purely thermal implementation, a temperature boost $\Delta T_{\rm SF}$ is defined and the heating of neighbouring gas particles is sampled stochastically given the energy available for feedback. A value of $\Delta T_{\rm SF}=10^{7.5}$K is chosen to be high enough to mitigate rapid cooling due to numerical effects, but low enough to avoid poor sampling of heating events around individual star particles \citep{DallaVecchia12}.  
\item {\em Seeding, merging, accretion and feedback from supermassive black holes} is implemented as described in \cite{Schaye15}. Briefly, dark matter halos with virial mass $>10^{10}\,h^{-1}{\rm M}_\odot$ are seeded with a black hole of mass $10^5\,h^{-1}{\rm M}_\odot$. These can grow through Eddington-limited accretion of gas while accounting for the gas angular momentum as by \citet{RosasGuevara15}, and through mergers with other black holes, following \cite{Springel05b} and \cite{Booth09}. Feedback from accreting black holes is also modelled by heating surrounding gas using an implementation similar to that of stellar feedback. The temperature boost for AGN heating events is chosen to be $T_{\rm AGN} = 10^{8.5}$K for all the simulations considered here. 
\end{itemize}

Each star particle represents a \lq simple stellar population\rq\ (SSP), characterised by an assumed stellar initial mass function (IMF, \eagle{} adapts the \cite{Chabrier03} IMF over the mass range [0.1,100]~M$_\odot$), and assuming that stars have a metallicity inherited from the converted gas particle, with a single age corresponding to the time that the gas particle was converted to a star particle. We then use published stellar life-times, evolutionary tracks, and yields to compute the rate at which these stars evolve and lose mass, as well as the rate of core collapse and Type Ia supernova events as described in \cite{Wiersma09b}. The simulation tracks 11 elements (H, He, C, Ni, O, Ne, Mg, Si, S, Ca, Fe) as well as a \lq total metallicity\rq\ (metal mass fraction) variable for each gas and star particle. The mass, age, and metallicity of the SSP are input parameters for the population synthesis model described below.

\begin{itemize}
\item {\em Dark matter halos} are identified using the \lq friends-of-friends\rq\ algorithm ({\sc fof}), linking dark matter particles within 0.2 times the mean inter-particle separation into a single {\sc fof} halo. Other particles are assigned to the same halo (if any) as the nearest dark matter particle. We characterise the mass of the halo by its $M_{200,{\rm crit}}$ value. This is the mass enclosed within a sphere of radius $R_{200,{\rm crit}}$ centred on the location of the particle with minimum gravitational potential in the halo. This radius is chosen such that the mean density within this sphere is 200 times the critical density, given the assumed cosmology.
\item {\em Galaxies} are identified with the {\sc subfind} algorithm \citep{Springel01,Dolag09}.
{\sc subfind} identifies self-bound substructures within halos which we associate with galaxies. The \lq central\rq\ galaxy is the galaxy closest to the centre of the parent {\sc fof} halo; this is nearly always also the most massive galaxy in that halo. The other galaxies in the same halo are its satellites. Particles in a halo not associated with a bound substructure ({\em i.e.} satellites) are assigned to the central galaxy. Central massive galaxies ($M_\star\ge 10^{11}~{\rm M}_\odot$, say) then have an extended halo of stars around them, usually referred to as intra-group or intra-cluster light. Determining the mass or indeed luminosity of such a large galaxy is ambiguous, both in simulations and in observations. For this reason we impose an aperture on the definition of a galaxy: we follow \citet{Schaye15} and calculate masses and luminosities for every subhalo, excluding material that is outside a 30~pkpc spherical aperture centred on the subhalo potential minima as well as material that is not bound to that subhalo. The 30~pkpc aperture has been shown to mimic an observational Petrosian aperture, and reduces intra-cluster light in massive centrals while lower mass galaxies are unaffected \citep{Schaye15}.
\end{itemize}

\subsection{Galaxy colours} 
The stellar population properties (age, metallicity \& assumed IMF) of an {\sc eagle} galaxy are combined with the \cite{bc03} population synthesis model to construct an SED for each star particle. Summing spectra over all stars within the aperture described in section \ref{sec:subgrid} and convolving with a filter response function yields broad-band colours, which we compute using
the {\it ugrizYJHK} photometric system for optical and near infrared photometry \citep[taken from][]{SDSSfilters, UKIRTfilters}. We express these absolute magnitudes in the AB-system, see \cite{Trayford15} for more details.

It is well known that dust can alter the optical colour of a galaxy significantly, particularly for gas-rich discs seen edge-on. We describe a simple model for dust reddening in a previous study \citep{Trayford15}, as well as a model that uses ray-tracing to account for the patchy nature of dust clouds enshrouding star-forming regions described in a forthcoming study (Trayford et al., {\em in prep.}). However, here we use the \lq intrinsic\rq\ ({\em i.e.} rest-frame and dust-free) colours of galaxies to examine the changes arising purely from the evolution of their stellar content. To simplify the interpretation we always quote {\em rest-frame} colours: there is therefore no \lq k\rq-correction needed to compare galaxies in the same band at different redshifts.  We concentrate here on {\it u$^{\star}$-r${^\star}$} colours (with the $^\star$ referring to intrinsic colours) rather than {\it g$^{\star}$-r${^\star}$}, because the {\it u} band is more sensitive to recent star formation, leading to more clearly separated blue/red colour sequences. Indeed, the {\it u$^{\star}$-r${^\star}$} index traverses the 4000\AA{} break, often used as a proxy for star formation activity \citep[e.g.][]{Kauffmann03}. The photometry is presented here without dust effects, comparison is possible with various observational data where dust corrections have been estimated \citep[e.g.][]{Schawinski14}.

%% file: Evolution.tex
\section{Colour evolution of the ensemble galaxy population}
\label{sec:colev}

\begin{figure*}
 \includegraphics[width=0.99\textwidth]{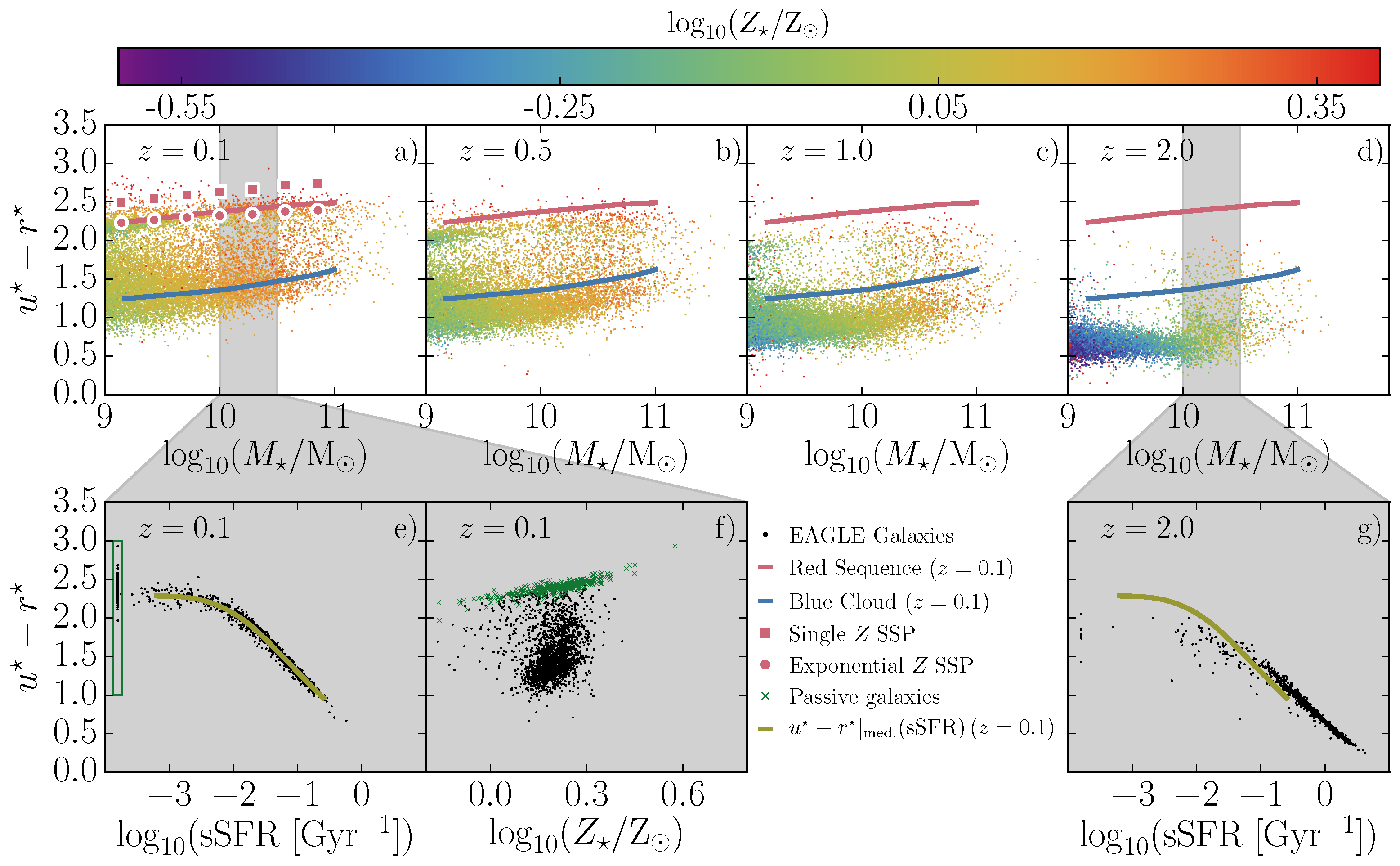}  

 \caption{Colour evolution of \eagle\ galaxies. {\bf Top row:} \us-\rs{} vs $M_\star$ colour-mass diagram at four redshifts ($z=0.1$, 0.5, 1 and 2, left to right). Individual galaxies are plotted as points, coloured by median stellar metallicity, using the colour bar in the bottom row. The locations of the red sequence and blue cloud at $z=0.1$ (red and blue lines, respectively) are repeated in panels b-d to guide the eye. Filled red squares show \us-\rs versus $M_\star$ for a 10~Gyr old stellar population with metallicity $Z_\star$ equal to the median metallicity at that $M_\star$; filled circles are the same, but assuming an exponential distribution of stellar metallicities with the same median. {\bf Bottom row, panels e and g:} dependence of \us-\rs colour on specific star formation rate (sSFR, $\dot M_\star/M_\star$) for galaxies with $10<\log_{10}(M_\star/{\rm M}_\odot)<10.5$ (the grey band in panel a, and galaxies with $M_\star$ between the two grey lines in panel d) at redshift $z=0.1$ and $z=2$, respectively. The olive line indicates the median \us-\rs{} as a function of sSFR at $z=0.1$ for comparison at $z=2$. {\bf Panel f:} \us-\rs\ versus median stellar metallicity for the galaxies of panel e; galaxies with sSFR$<10^{-3.5}$~Gyr$^{-1}$, appearing in the green box in panel e, are plotted as green dots.}
 \label{fig:col}
 
\end{figure*}

\begin{figure*}
 \includegraphics[width=0.99\textwidth]{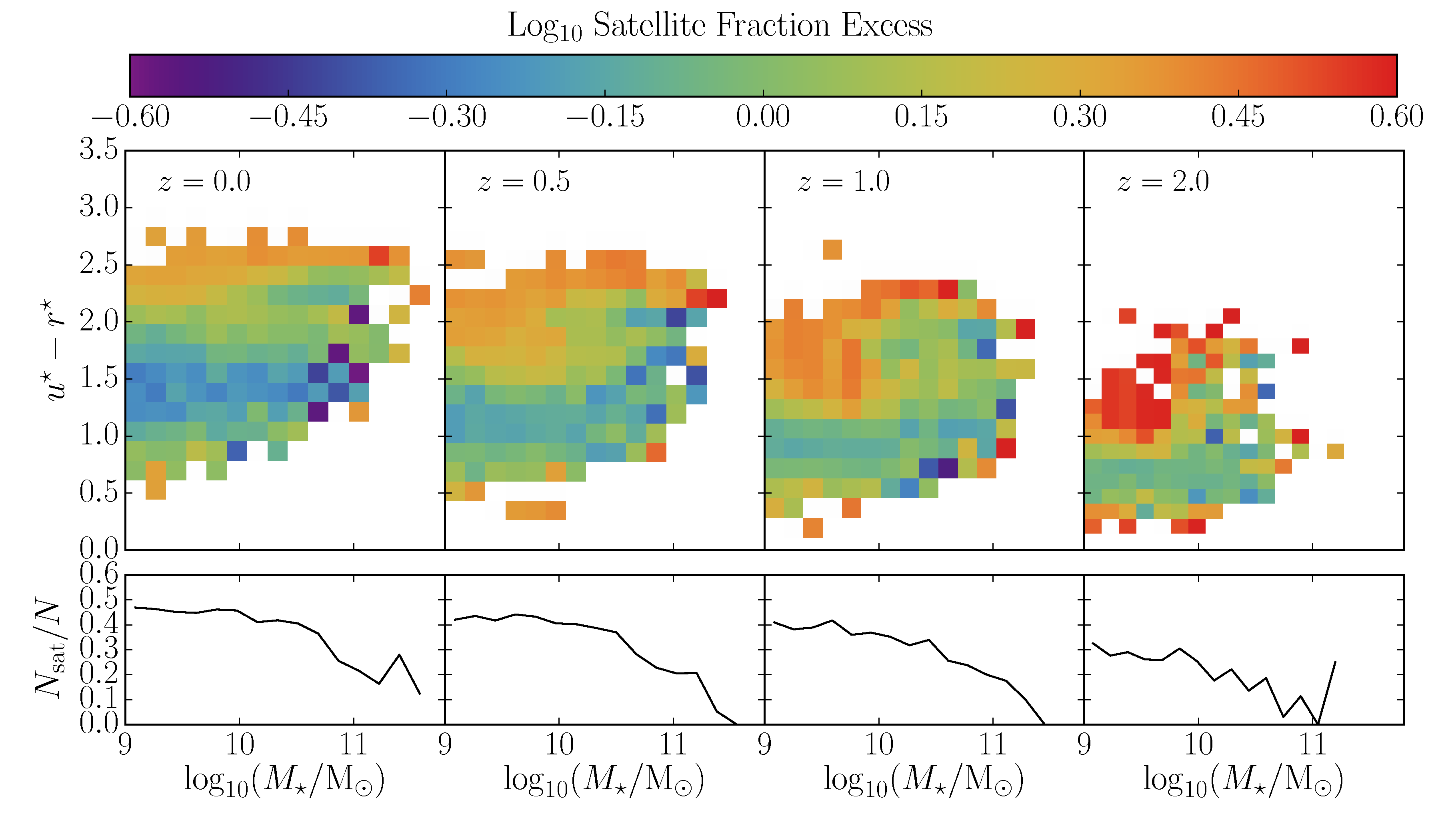}  

 \caption{The impact of satellite fraction on the evolution of the \us-\rs vs $M_\star$ colour-stellar mass relation.
 {\bf Top panels:} Each square corresponds to a bin in colour and $M_\star$ and is coloured according to the median normalised satellite fraction in that bin, such that higher satellite fractions correspond to redder colours (see the colour bar).  The satellite fraction is normalised to the average satellite fraction at that stellar mass (bottom panel), removing trends of satellite fraction with stellar mass and redshift.
 At $z=2$, most red galaxies with $M_\star\le 10^{10}\,{\rm M}_\odot$ are satellites (red colour in satellite fraction). This trend persists to $z=0$, although it becomes weaker as galaxies classified as centrals also get quenched. {\bf Bottom panels:} Fraction of galaxies classified as satellites as a function of $M_\star$. At redshift $z=0$, the satellite fraction is nearly constant at just below 50 per cent below $M_\star=10^{10}{\rm M}_\odot$, and decreases above that mass. At higher $z$ the satellite fraction decreases for all $M_\star$.}
 \label{fig:sat}
 
\end{figure*}

\begin{figure*}
 \includegraphics[width=0.99\textwidth]{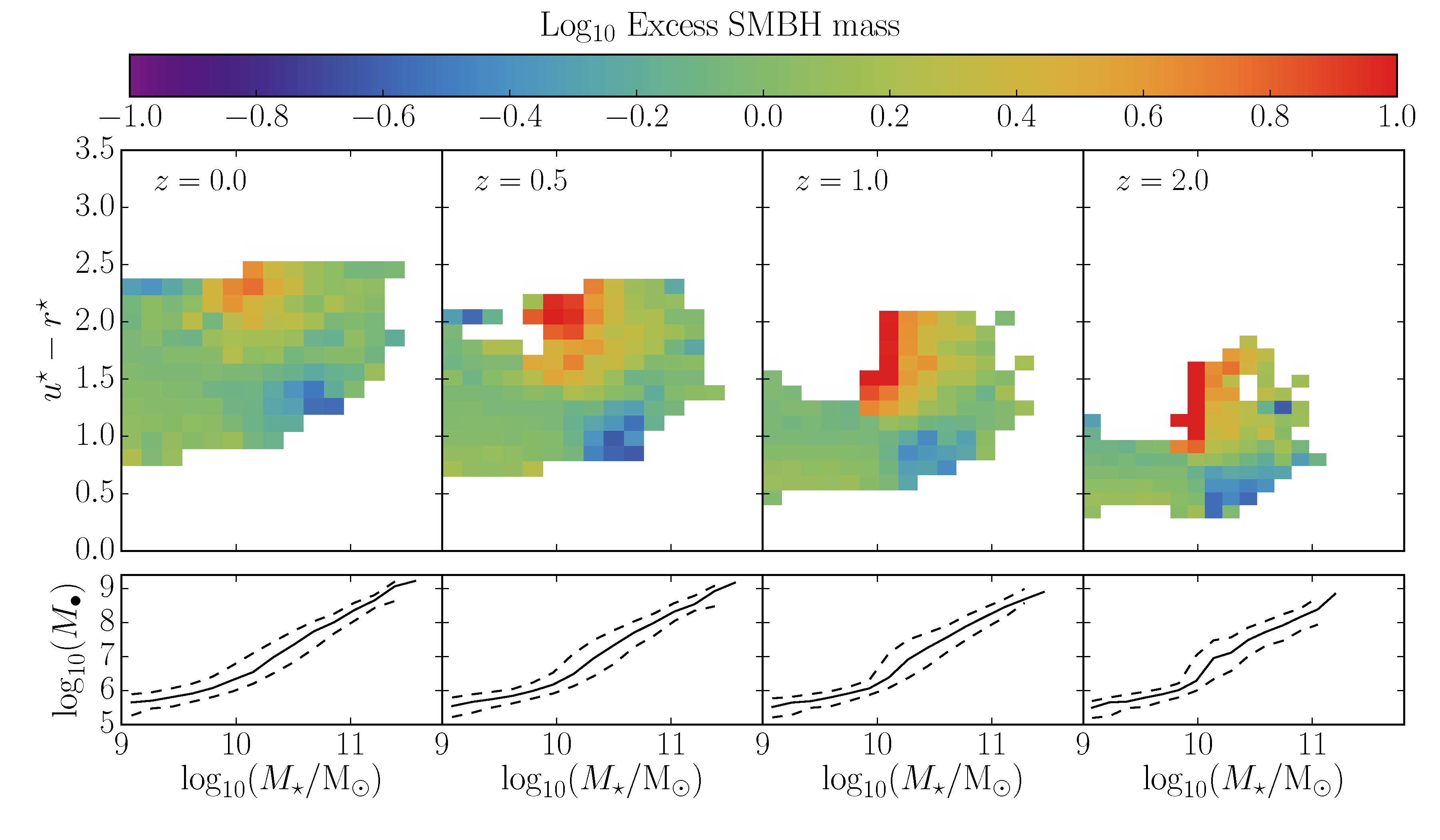}  

 \caption{The impact of black hole mass on the evolution of the \us-\rs{} vs $M_\star$ colour-stellar mass relation for central galaxies. {\bf Top panels:} Each square corresponds to a bin in \us-\rs{} and $M_\star$, and is coloured according to the median black hole mass, $M_\bullet$, in that bin, such that larger values of $M_\bullet$ correspond to redder colours (see the colour bar). The median black hole mass in each square is normalised to the median black hole at that stellar mass (bottom panel), removing trends of $M_\bullet$ with $M_\star$ and redshift. At $z=2$ there is a trend for redder galaxies to have more massive black holes. This trend is particularly striking for galaxies with $M_\star\sim 10^{10}{\rm M}_\odot$ and becomes less pronounced at higher masses. There is no obvious correlation at lower stellar mass. These trends persist to lower $z$ but become weaker. {\bf Lower panels:} The median black hole mass, $M_\bullet$, as a function of stellar mass is plotted in solid black. Dashed black lines represent the 16th and 84th percentiles. $M_\bullet$ is nearly independent of $M_\star$ below $M_\star\sim 10^{10}{\rm M}_\odot$, and increases with $M_\star$ above this characteristic mass. This trend is almost independent of redshift.
}   
\label{fig:bh}
\end{figure*}

Figure~\ref{fig:col}a shows that a scatter plot of \eagle{} galaxies in a colour-stellar mass diagram, (\us-\rs{}) vs $M_\star$, exhibits strong bimodality in colour at redshift $z\approx 0$. The well defined red sequence resides at \us-\rs$\gtrsim 2.2$ with colours becoming redder with increasing $M_\star$. The blue cloud is at $u^\star-r^\star\approx 1.3$, with a slope similar to that of the red sequence. These two sequences are indicated by red and blue lines to guide the eye, respectively, obtained by a spline fit to the maxima in the probability distribution of \us-\rs\ in bins of $M_\star$. We keep the location of these lines fixed in Fig.~\ref{fig:col}b-d to facilitate comparison at higher $z$. We clearly see that:
\begin{itemize}
\item[({\em i})] The red sequence becomes bluer and less populated with increasing $z$. It is in place at $z\approx 1$ but has mostly disappeared by $z\approx 2$. A gap in the red sequence is noticeable at $z\approx 1$ for $M_\star \sim 10^{9.7}$~${\rm M_\odot}$.
\item[({\em ii})] The blue sequence becomes bluer with increasing $z$, and exhibits decreasing scatter.
\end{itemize}

The main features of the galaxy population that drive these trends are illustrated in the bottom panels of the figure. Figs.~\ref{fig:col}e \& g show that for a narrow stellar mass range around $M_\star=10^{10.25}~{\rm M}_\odot$ that \us-\rs\ is strongly anti-correlated with the specific star formation rate, ${\rm sSFR}\equiv \dot M_\star/M_\star$, provided the galaxy is star-forming ($\log_{10}({\rm sSFR}/{\rm Gyr})\gtrsim -2$). This is not surprising since the light in the \us\ filter is dominated by emission from massive and hence young stars, while \rs\ is dominated by the older population. Galaxies in this plot follow a very tight relation at a given $z$, sliding along a narrow locus in colour that becomes bluer at higher $z$. At $z=2$, the galaxies follow almost the same relation in \us-\rs\ versus sSFR as at $z=0$, with just a small but noticeable offset towards redder colours ($\approx 0.1$ mag) for $\log_{10}({\rm sSFR}/{\rm Gyr})\gtrsim -1.25$. This is a result of the redder population being younger on average and hence brighter for a $z=2$ star-forming galaxy, compared to a star-forming galaxy at $z=0$. For lower star formation rates ($\log_{10}({\rm sSFR}/{\rm Gyr})\lesssim -1.25$) the old population has more influence on \us-\rs{}, and the younger average stellar age of $z=2$ galaxies causes an offset to blue colours. 

With \us-\rs\ colour so strongly correlated with sSFR, the colour vs $M_\star$ diagram of Fig.~\ref{fig:col} is one view of the
	\lq fundamental plane of star-forming galaxies\rq, discussed recently by \cite{Lagos15b}. These authors showed that \eagle\ galaxies from different redshifts fall onto a single 2D surface when plotted in the 3D space of $\dot M_\star - M_\star$ and gas fraction (or metallicity), which they attributed to self-regulation of star formation. \cite{Lagos15b} also showed that observed galaxies follow very similar trends. The increasingly bluer colours of the blue cloud towards higher $z$ is a consequence of the increased star formation activity at fixed $M_\star$.

The scatter in colour at fixed $M_\star$ on the red sequence is mostly due to metallicity, $Z$,
 as is clear from examination of the \us-\rs\ distribution of galaxies at a given $M_\star$ with low sSFR $<10^{-3}$~Gyr$^{-2}$, plotted as green points in Fig.~\ref{fig:col}f. The colour of star-forming galaxies with sSFR $>10^{-3}$~Gyr$^{-1}$ (black points)  also depends on $Z$, but from comparison of these panels it is clear that this effect is much smaller than the dependence of colour on sSFR itself - it induces the small scatter in \us-\rs\ in panel e, on top of the main trend with sSFR.

As discussed by many others, the {\em slope} of the red sequence demonstrates the dependence of colour on $Z$ for galaxies: more massive galaxies are more metal rich and hence redder (see \citealt{Trayford15}). Because the mass- and light-weighted metallicities are not equivalent, the internal metallicity distribution for stellar populations in a galaxy may also affect the normalisation of the red sequence. To illustrate this, we calculated the median metallicity, $Z_{\rm med}(M_\star)$, in bins of stellar mass. We then calculated \us-\rs\ colours for a 10~Gyr old population with that dependence of $Z$ on $M_\star$, and plot the resulting \us-\rs\ colour as a function of $M_\star$ in Fig.~\ref{fig:col}a as red squares. Although this sequence has the same slope as the red sequence in \eagle, it is systematically redder by $\approx 0.25$~magnitudes. This is not an age effect, but a consequence of stellar populations exhibiting a spread in metallicity within an \eagle\ galaxy. In fact, the metallicity distribution function of stars in an \eagle{} galaxy is fairly well described by an exponential distribution. We therefore generated another comparison toy model for the red sequence colour, in which we impose an exponential metallicity distribution and again assume a coeval 10 Gyr old population. The exponential metallicity distribution is defined by a mean value at fixed mass, given by the $Z_{\rm med}(M_\star)$ dependence of \eagle{} galaxies. This model is plotted as filled red circles and it reproduces the \eagle{} red sequence very well. This simple exercise shows that the assumption that all stars have the {\em same} metallicity results in systematic errors in the metallicity from broad-band colours.

The consistent red sequence slope between the toy model and \eagle{} suggests that any changes in the internal stellar $Z$ distribution of \eagle{} galaxies with mass are not strong enough to bias the median colours of red galaxies. We note that the slope of $u^\star-r^\star$ relation as a function of $M_\star$ in the blue cloud is set by the sSFR-$M_\star$ relation and not by metallicity effects. Therefore, the similarity between the slopes of the blue and red lines is coincidental.

\subsection{Satellite colours}
\label{sec:satellites}

The extent to which satellite galaxies are preferentially red relative to the general population is illustrated in Fig.~\ref{fig:sat}. We divide colour vs mass diagrams at different redshifts into equal bins of \us-\rs{} and $\log_{10}(M_\star/{\rm M}_\odot)$. The satellite fraction in each bin is computed and normalised by the total satellite fraction for all galaxies in the same stellar mass range. Bins containing $> 10$ galaxies are shaded by the $\log_{10}$ normalised satellite fraction, such that positive values indicate a higher than average satellite fraction for that mass, while negative values indicate a lower than average value. The satellite fraction as a function of stellar mass in \eagle{} is plotted for each redshift in the bottom panels.

Galaxies with $M_\star\le 10^{10}{\rm M}_\odot$ that are red are predominately satellites, seen most strikingly at $z\approx 2$. At lower $z$ there is still a trend for low-mass red galaxies to be satellites, but the trend is less pronounced because some galaxies classified as centrals are also red. To some extent this may be a consequence of galaxies being quenched by ram-pressure stripping in the outskirts of more massive halos, {\em before} they are classified as being a satellite \citep[e.g.][]{Bahe13}. Indeed, they may not be part of the {\sc fof} halo (yet). Another possibility is that some of these galaxies were stripped as satellites when they fell inside a massive halo but have travelled out again, the so-called backsplash population \citep{Balogh00}.

At redshift $z\approx 0$, the fraction of satellites is $\approx 50$~per cent at $M_\star\sim 10^9{\rm M}_\odot$, decreasing slowly to 30 per cent by $M_\star\sim 10^{10.5}{\rm M}_\odot$, and then dropping rapidly towards higher $M_\star$. Satellite fractions decrease slowly at all $M_\star$ with increasing $z$ to $z\approx 1$, and then drop much faster to below 30 per cent at all masses by $z=2$. This rapid drop in the satellite fraction with increasing $z$ is the reason that the red sequence disappears at low $M_\star\lessapprox 10^{10}{\rm M}_\odot$ for $z\gtrsim 2$.

\subsection{AGN host colours}
\label{sec:agn}

The effect of feedback from accreting black holes on galaxy colours is illustrated in Fig.~\ref{fig:bh}. We only plot {\em central} galaxies to disentangle satellite quenching from effects induced by AGN. The figure is analogous to Fig.~\ref{fig:sat}, with median black hole mass replacing satellite fraction. The median black hole mass ($M_\bullet$) as a function of $M_\star$ is  plotted for each redshift as the bottom panels in Fig.~\ref{fig:bh}.

At redshift $z=2$, there is a very strong trend for red galaxies with $M_\star\approx 10^{10}$~${\rm M}_\odot$ to exhibit unusually high black hole masses, more than 4 times the median black hole mass at that stellar mass. This trend persists but becomes weaker at higher $M_\star$, and is completely absent at lower masses. This correlation between the residuals of the $M_\bullet-M_\star$ relation and the colour of the galaxy is still mostly present at $z=1$, but begins to be washed out at later times.

The trend for galaxies with high black hole masses to be predominantly red when $M_\star\gtrapprox 10^{10}$~${\rm M}_\odot$ is likely related to the largely redshift independent characteristic {\em halo} mass, $M_{\rm h} \sim 10^{12}$ M$_\odot$, above which black holes start to grow rapidly in \eagle{}, fed by the growing hot halos around them (Bower et al. {\em in prep.}). The accreting black hole then quenches star formation in its host galaxy, turning it red. A corollary of the existence of this characteristic halo mass, is that black holes only start to grow significantly when the galaxy's stellar mass is $\sim 10^{10}$~${\rm M}_\odot$ (Fig.~\ref{fig:bh}, bottom panels). The scatter in the $M_\star-M_{\rm h}$ relation results in the transition between dormant and rapidly growing black holes being less well-defined in the $M_\bullet-M_\star$ relation in comparison to a $M_\bullet-M_{\rm h}$ plot.

\subsection{Colour transformation mechanisms}
\label{sec:coltrans}

Combining the results of Figs.~\ref{fig:sat} and ~\ref{fig:bh} enables us to understand the origin of the evolution in the \us-\rs{} vs $M_\star$ diagram of Fig.~\ref{fig:col}: galaxies with $M_\star< 10^{10}$~${\rm M}_\odot$ tend to become red when they become satellites, whereas galaxies above this characteristic mass are quenched by their AGN. This reasoning also explains why the red sequence starts to build-up from both the low-mass and the high-mass ends, leaving initially a noticeable scarcity of red galaxies at $M_\star\approx 10^{9.7}{\rm M}_\odot$ at $z\approx 1$. Such galaxies are too low mass to host a vigorously accreting black hole, yet too massive to be satellites in the typically lower-mass groups at that higher $z$. It is not until redshifts $z<1$ that the more massive halos that host $M_\star\approx 10^{10}$~${\rm M}_\odot$ satellites appear.
	
The extent to which \eagle{} predicts the characteristic stellar mass above which AGN quenching occurs, and the evolution of the under-abundance of intermediate-mass red galaxies, not only depends on the details of the subgrid physics but also on the volume that is simulated. This is because massive clusters are under-represented or simply absent due to missing large-scale power in the density field, and poor sampling of rare objects in the relatively small, periodic \eagle\ volume of 100$^3$~cMpc$^3$. However we believe that the relevant physics described here is robust, and corroborates the similar conclusions of \cite{Gabor12} who used an \textit{ad-hoc} model for quenching in massive galaxies, as opposed to our physically motivated subgrid scheme for the \eagle{} simulations that are implemented on smaller (sub-kpc) scales.

A corollary of satellite quenching for lower-mass galaxies, and AGN quenching for more massive galaxies, is that these low- and high-mass red galaxies tend to inhabit the {\em same} dark matter halos. The more massive red galaxy is the central galaxy of this halo and is quenched by its AGN. Conversely, the lower-mass red galaxies are the satellites of a massive central red galaxy.  As a consequence, the low- and high-mass red galaxies have similar clustering strengths, with both clustering more strongly than blue galaxies. \eagle{} reproduces the observed clustering as a function of colour and luminosity well, as will be discussed by Artale et al. ({\rm in prep.}). We  next investigate how, and at what rate, individual galaxies move through the \us-\rs{} vs $M_\star$ diagram.

%% file: Quenching.tex
\section{Colour evolution of individual galaxies}
\label{sec:quench}

\begin{figure*}
 \includegraphics[width=0.98\textwidth]{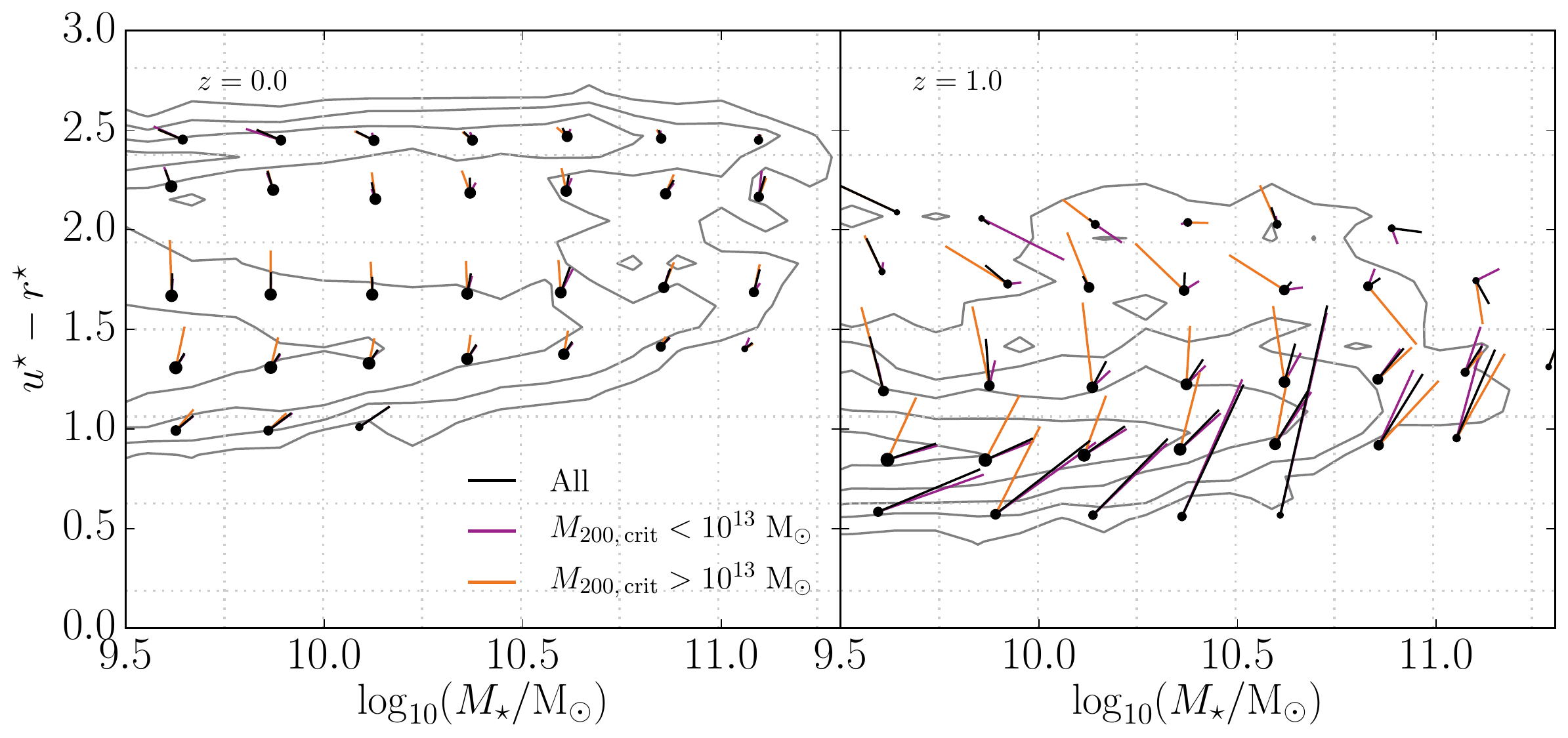}  

 \caption{The flow of galaxies in (\us-\rs, $M_\star$) space between two redshifts, $z_1$ to $z_2$. \textit{Right panel} shows the galaxy flow between $z_1=1.3$ and $z_2=1$ snapshots, a period of $\approx 0.9$ Gyr. The \textit{left panel} shows the galaxy flow interpolated between the $z=0.1$ and $z=0$ snapshots to yield the same time period, with $z_1 = 0.07$ and $z_2 = 0$. {\em Black circles} represent the mean location of galaxies at $z_1$, selected in a bin of \us-\rs - $M_\star$; the size of the circle is proportional to the logarithm of the total stellar mass in galaxies in that bin. {\em Black vectors} represent the mean motion of the galaxies in that bin between $z_1$ and $z_2$. {\em Orange vectors} (purple vectors) are for those galaxies that at redshift $z_2$ belong to halos with virial mass $M_{\rm 200, \, crit}>10^{13}$M$_\odot$ ($M_{\rm 200, \, crit}<10^{13}$M$_\odot$). Centre coordinates and vectors sampling fewer than $10$ galaxies are not plotted. The overall distribution at the later redshift is plotted as \textit{grey contours} for comparison. We do not take into account galaxies merging into hosts that are more than four times their mass, illustrating such mergers in more detail in Fig. \ref{fig:massive_track}.}

\label{fig:massflow}
\end{figure*}

\subsection{The flow of galaxies in the colour-$M_\star$ plane}
\label{sec:massflow}

Fig.~\ref{fig:massflow} illustrates how galaxies move through the (\us-\rs, $M_\star$) plane. Selecting galaxies in equal bins of \us{}-\rs{} and $\log_{10} (M_\star)$ at one redshift, we measure the median difference in \us{}-\rs{} and $\log_{10} (M_\star)$ for their descendant galaxies at a second redshift. We plot these differences for galaxies over an equal time period at high ($z\approx1$) and low ($z\approx0$) redshift. This is achieved by using two consecutive snapshots ($z=1.3$ and $z=1$) for the right panel, corresponding to a time interval of $\approx0.9$ Gyr, and interpolating the galaxy vectors between the lowest redshift snapshots ($z=0.1$ and $z=0$) to match the same time period. Descendants that grow in $M_\star$ by a factor $>4$ through merging into a more massive host galaxy are eliminated from the measurement, to prevent them contributing extreme vectors to their bin.

For both redshift ranges, it is clear that the colours of galaxies generally become redder, with vectors pointing in the positive \us{}-\rs{} direction. Exceptions can be seen on the red sequence, in which some red galaxies become star-forming following a gas-rich merger - we discuss the fraction of such \lq rejuvenated\rq\ galaxies below. Red sequence galaxies show little change in \us-\rs{}, but in general those with $M_\star \lesssim 10^{10.75}{\rm M_\odot}$ lose mass, with only the most massive red sequence galaxies showing mass growth. Considering that we do not count mergers into hosts of factor $>4$ higher $M_\star$, this suggests that red sequence galaxies are being stripped prior to a dry merger with a massive central. We also see evidence of \lq mass quenching\rq, with the vectors for blue-cloud selected galaxies becoming steeper with increasing stellar mass. For $M_\star > 10^{10}{\rm M_\odot}$ this is attributable to the presence of AGN (Fig.~\ref{fig:bh}).

Another notable behaviour seen in Fig.~\ref{fig:massflow} is that the bluest galaxies (\us{}-\rs{} $\approx 0.5$, right panel) tend to change their \us-\rs colour {\em more} than the average blue galaxy, to the extent that they end-up on the {\em red} side of the blue sequence at the later redshift. The strong reddening and mass increase of these galaxies suggests that they are starbursts triggered just prior to a merger. Such a scenario is consistent with the fact that a higher proportion of satellites are found at these colours than at the centre of the blue peak (see Fig.~\ref{fig:sat}). Note that this effect is only observed for the high redshift panel, partially due to the bluest bins failing to meet the minimum galaxy count criterion of 10. These extreme starbursts are clearly rarer at low redshift.     

For each redshift range, we see that galaxies in more massive halos (orange vectors) have a stronger median shift in \us{}-\rs{} than their low-mass halo counterparts (purple vectors), suggesting that they have a higher likelihood of quenching.Galaxies in massive halos also generally exhibit more mass loss than the overall population, showing the role of environment in how galaxies evolve in the (\us-\rs, $M_\star$) plane.

\subsection{Evolution of colour populations in {\sc eagle}}
\label{sec:time}
\begin{figure*}
\includegraphics[width=0.99\textwidth]{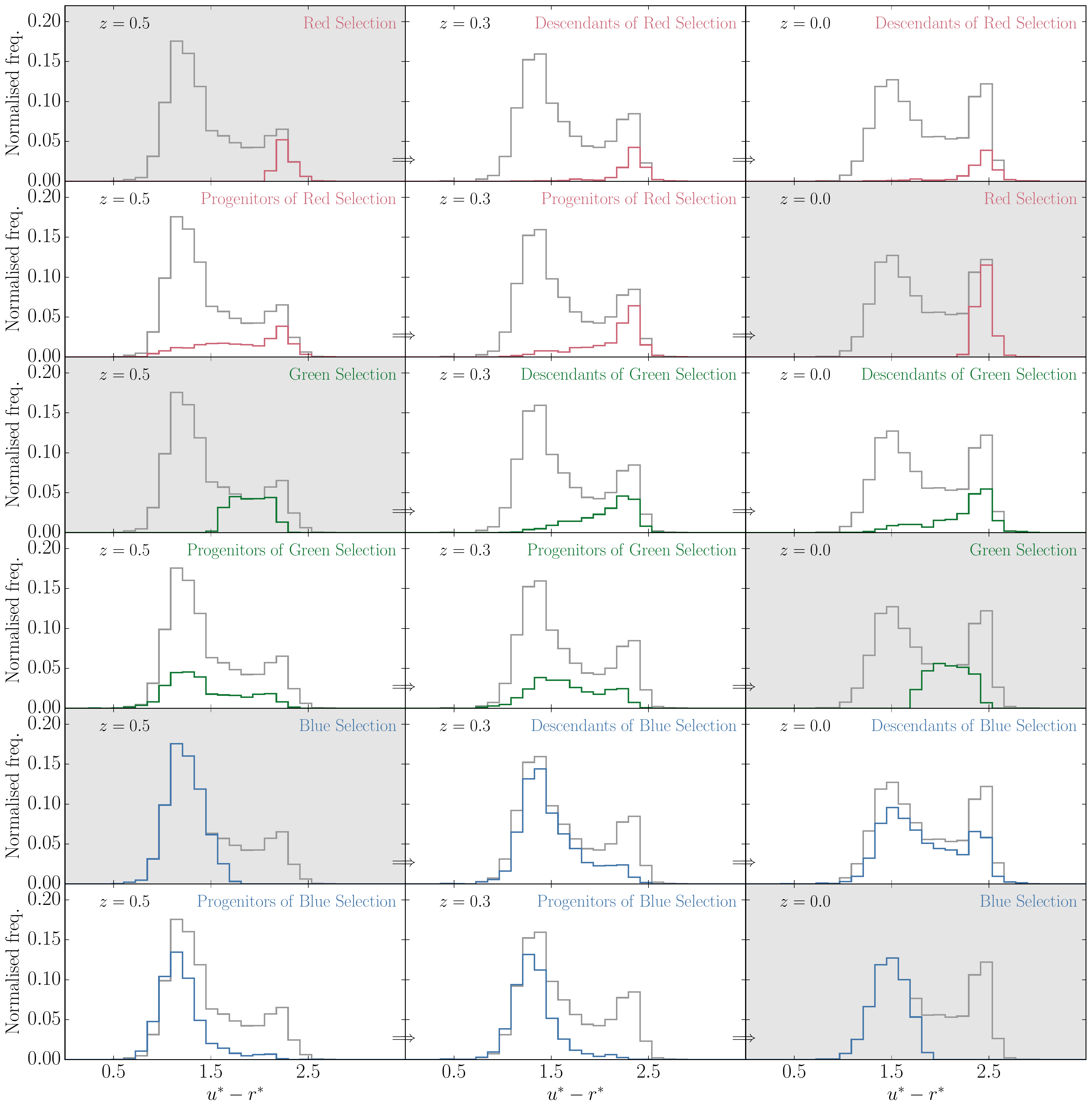}  

 \caption{{\em Grey histograms:} \us-\rs\ colour evolution of all galaxies from redshift $z=0.5$ to $z=0$, with redshift decreasing from \textit{left} to \textit{right} in each row; each panel is labelled with the corresponding redshift. Only galaxies with $M_\star > 10^{10}$M$_\odot$ are included. Colour selections are made using Eq.~\ref{eq:select} and are as follows: {\em top row:} we select red galaxies at $z=0.5$ (red histogram) and  plot the colour distribution of their descendants at low $z$ as a red histogram. {\em Second row from top:} we select red galaxies at $z=0$ (red histogram), and plot the colour distribution of their main progenitors as a red histogram at higher $z$. {\em Rows 3 and four from the top:} as above, but for green galaxies. {\em Bottom two rows:} as above, but for blue galaxies. The background colour of the panel in which galaxies were selected is coloured grey for ease of reference. }
 \label{fig:hist}
 
\end{figure*}

\begin{figure*}
 \includegraphics[width=0.99\textwidth, clip]{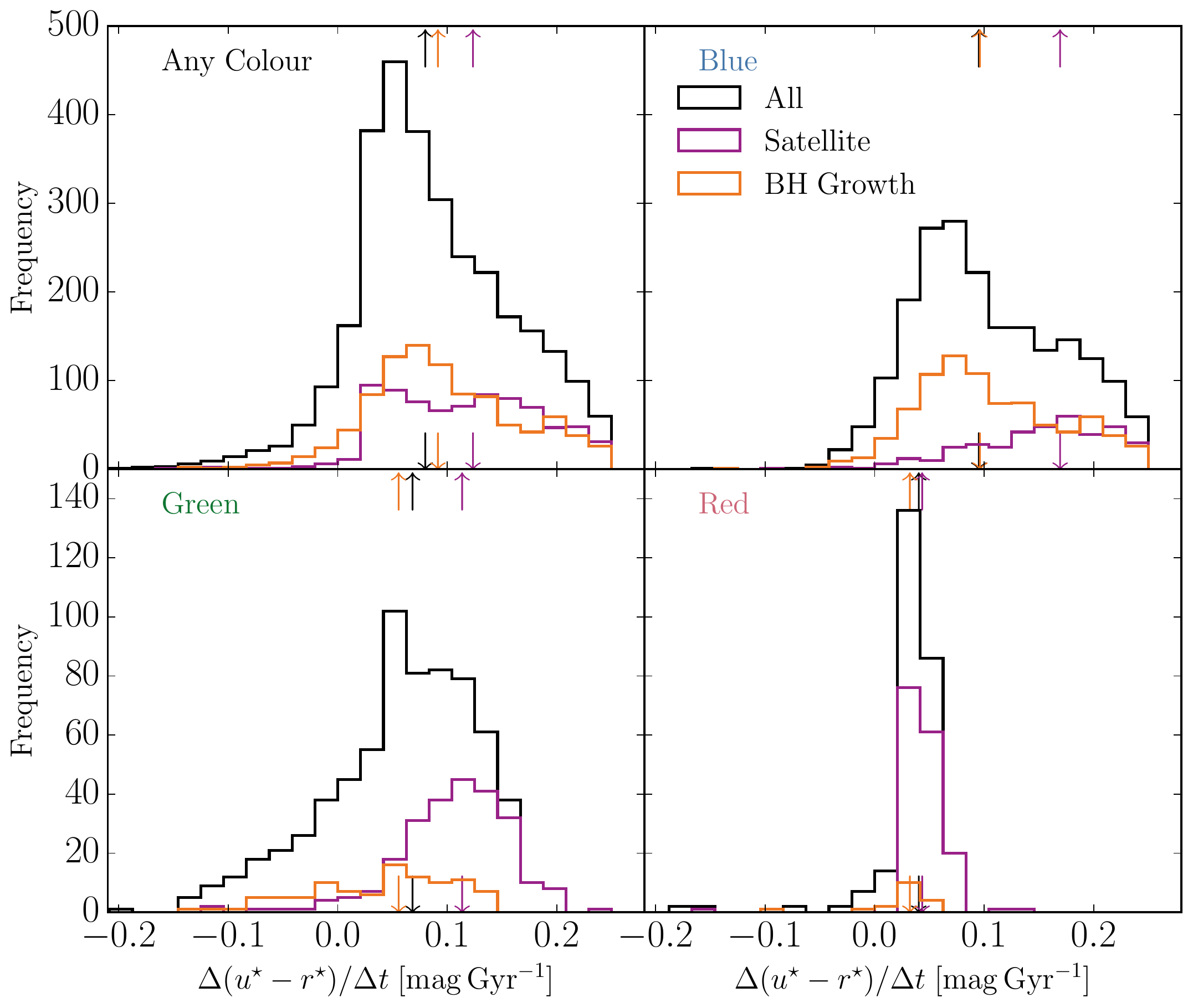}  

 \caption{{\bf Top left panel:} Distribution of the average rate of colour change over the redshift interval  $z=0.5\rightarrow 0$, $\Delta(\hbox{\us-\rs})/\Delta t$, for galaxies with mass $M_\star > 10^{10}{\rm M_\odot}$; vertical arrows (top and bottom) denote the median rate. {\bf Other panels:} As the top left panel, but for galaxies selected at $z=0.5$ to be blue, red, and green (clock wise from top right). In each panel, {\em black lines} refer to all selected galaxies, {\em purple lines} to the fraction that at $z=0.5$ are satellites, {\em orange lines} to the fraction whose central black hole has grown by at least a factor of 1.5 between $z=0.5$ and $z=0$. The rate of change of the median colour is typically small, but individual galaxies can change colour more dramatically over this period, $-0.2<\Delta(\hbox{\us-\rs})<0.25$ up to $\vert \Delta(u^{\star} - r^{\star}) / \Delta t \vert = 0.25$ Gyr$^{-1}$, particularly for green and blue galaxies.}
  \label{fig:speed}
 \end{figure*}
 
To track the evolution of galaxies selected to be \textit{red}, \textit{blue} or \textit{green}, we must first define these populations. To do this we apply cuts that evolve with redshift $z$ for red and blue galaxies.
\begin{eqnarray} 	
\label{eq:select}
&(u^\star - r^\star)_{\rm red} & > 0.2\log_{10}(M_\star/{\rm M_\odot}) - 0.25\,z^{0.6} + 0.24 \nonumber\\
&(u^\star - r^\star)_{\rm blue} & < 0.2\log_{10}(M_\star/{\rm M_\odot}) - 0.25\, z^{0.6} - 0.3\,
\end{eqnarray}  
The green galaxies are taken to be those that are not included in either set. These cuts are defined in an \textit{ad-hoc} way to divide the galaxies into three populations at each redshift. This is a similar procedure to that used by many observational studies, with authors adopting differing functional forms and normalisations (see e.g. the discussion in \citealt{Taylor14}). The exact form of the colour cuts is unimportant for our qualitative analysis, but is considered when we discuss our quantitative results. 

The evolution of the \us-\rs\ colours of galaxies, selected by colour either at high redshift ($z=0.5$) or low redshift ($z=0.1$), is illustrated in Fig.~\ref{fig:hist}. We bin galaxies in the three colour bins described above and plot the colour distribution of the descendants and main progenitors (odd and even rows, respectively) of galaxies selected to be red, green or blue (top two, middle two, and bottom two rows, respectively). We use the galaxy merger trees to identify descendants and main progenitors. For each panel, the \us-\rs{} colour distribution of all galaxies at the indicated redshift, with $M_\star > 10^{10}{\rm M_\odot}$, is plotted in grey. 
	
From the top two rows it becomes clear that most galaxies that are red at $z=0.5$ stay red to $z=0$, whereas a substantial fraction of galaxies that are red at $z=0$ were green at $z=0.1$ or even blue at $z=0.5$. Galaxies that are green at $z=0.5$ predominantly become red at $z=0$, but a fraction of green galaxies becomes blue (third row). Galaxies that are green at $z=0$ had a range of colours at $z=0.5$, although they were bluer than average (fourth row). The similar and dominant blue (red) fractions in the two green progenitor (descendant) panels suggests that the typical time to transition through the green valley is shorter than the redshift intervals used here. Finally, galaxies that are blue at $z=0.5$ have a large range of colours at $z=0$ with a distribution that is similar to that of the population as a whole (fifth row), whereas galaxies that are blue at $z=0$ were mostly blue at $z=0.5$ as well (bottom row).

The rate at which galaxies with stellar mass $M_\star > 10^{10}{\rm M_\odot}$ at $z=0.5$ change \us-\rs\ colour over the redshift range $z=0.5$ to $z=0$ (elapsed time $\Delta t\approx5$~Gyr) is quantified in Fig.~\ref{fig:speed}. We identify the $z=0$ descendant for all galaxies with $M_\star > 10^{10}{\rm M_\odot}$ at $z=0.5$, compute the change in colour, $\Delta($\us-\rs), and plot a histogram of rates, $\Delta($\us-\rs$)/\Delta t$. We also identify if a galaxy is a satellite at $z=0.5$, or if the mass of its central black hole increases by a factor $\geq 1.5$. This threshold is chosen to represent an above average black hole growth, while still providing a significant sample of galaxies.

The rate of change of the {\em median} colour of galaxies is small, $\Delta(\hbox{\us-\rs})/\Delta t\approx 0.08$~mag~Gyr$^{-1}$ to the red, but is larger for galaxies whose black hole grows more than average ($\Delta(\hbox{\us-\rs})/\Delta t\approx 0.09$~mag~Gyr$^{-1}$) or those that are satellites ($\Delta(\hbox{\us-\rs})/\Delta t\approx 0.12$~mag~Gyr$^{-1}$). Galaxies that are red at $z=0.5$ typically change little in colour to $z=0$, ($\Delta(\hbox{\us-\rs})/\Delta t\approx 0.03$~mag~Gyr$^{-1}$), except for the occasional outlier that becomes blue. The rate of change of the median colour is larger for galaxies that are green or blue at $z=0.5$, with individual 
	galaxies changing colour more rapidly, both to the red and to the blue. Galaxies that are satellites can undergo rapid
	changes to the red, $\Delta(\hbox{\us-\rs})/\Delta t\gtrsim 0.2$~mag~Gyr$^{-1}$, whether blue or green at $z=0.5$. Note that this rate is averaged over a considerable period ($\approx$ 5 Gyr), and instantaneous rates of colour change for galaxies can be much higher, as explored below.   

\subsection{Colour-mass tracks of individual galaxies}
\label{sec:tracks}

\begin{figure*}
 \includegraphics[width=0.99\textwidth]{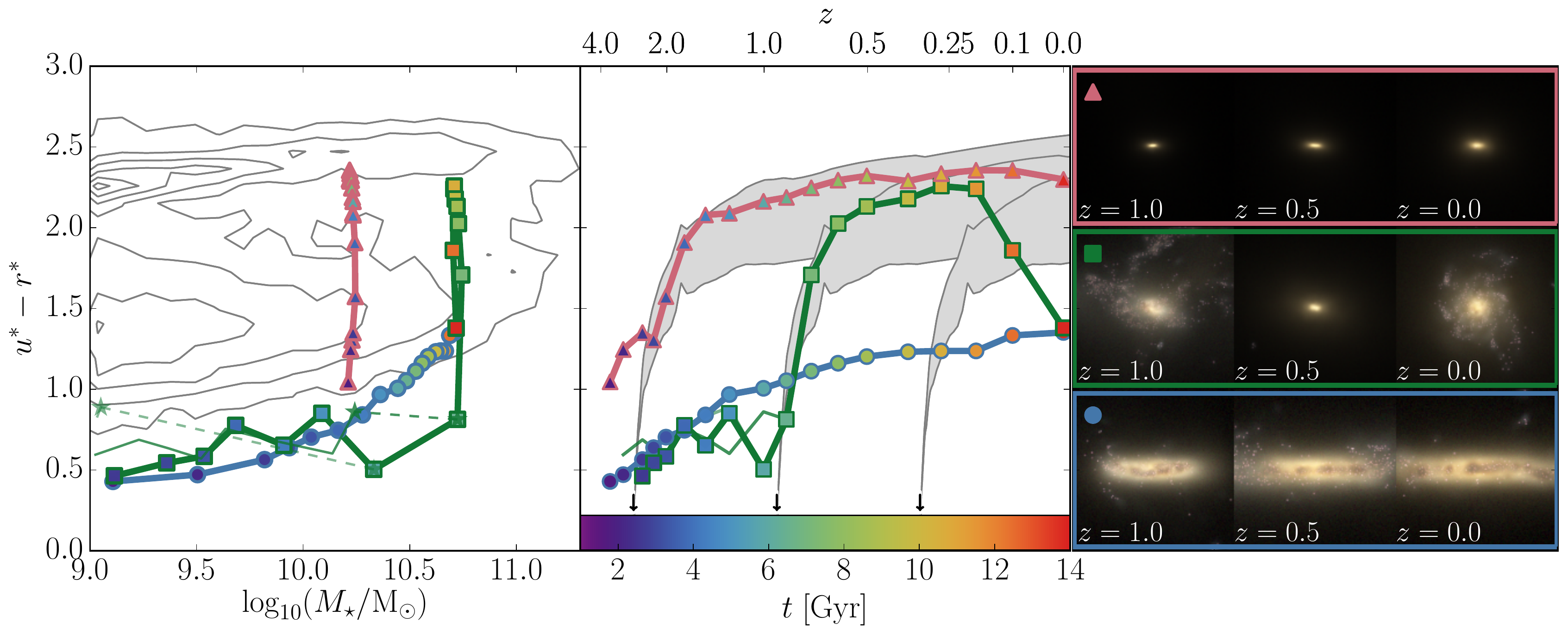}

 \caption{Tracks illustrating the change in mass, colour and morphology, of a quiescently star-forming galaxy (thick blue curve \& filled circles), a rapidly quenched galaxy (thick red curve \& filled triangles), and a rejuvenated galaxy (thick green curve \& filled squares). Thin tracks show merging satellites (see caption of Fig. \ref{fig:massive_track}). {\em Left and middle panels:} tracks in the (\us-\rs, $M_\star$) plane (left panel) and colour as function of time and redshift (middle panel), from redshift  $z=4$ to $z=0$. Symbol colour corresponds to cosmic time as per the colour bar.  Background contours in the left panel correspond to the $z=0$ colour-$M_\star$ distribution. Grey tracks in the middle panel depict the colour evolution of a passively-evolving coeval starburst (indicated with an arrow). Each burst is assumed to be composed of stars with an exponential distribution of metallicities with given mean. The width of the grey region corresponds to varying this mean metallicity over the range of [1/3,3] times solar ($Z_\odot=0.0127$). {\em Right panel:} edge-on gri-composite image of side length 40 pkpc, calculated using ray-tracing to account for dust (Trayford et al. 2015, {\em in prep.}), for the $z=0$ galaxy and its $z=0.5$ and $z=1$ main progenitor. The corresponding symbol for each track is indicated on galaxy images.}
 \label{fig:lowmass_track}
 \end{figure*}

\begin{figure*}
 \includegraphics[width=0.99\textwidth]{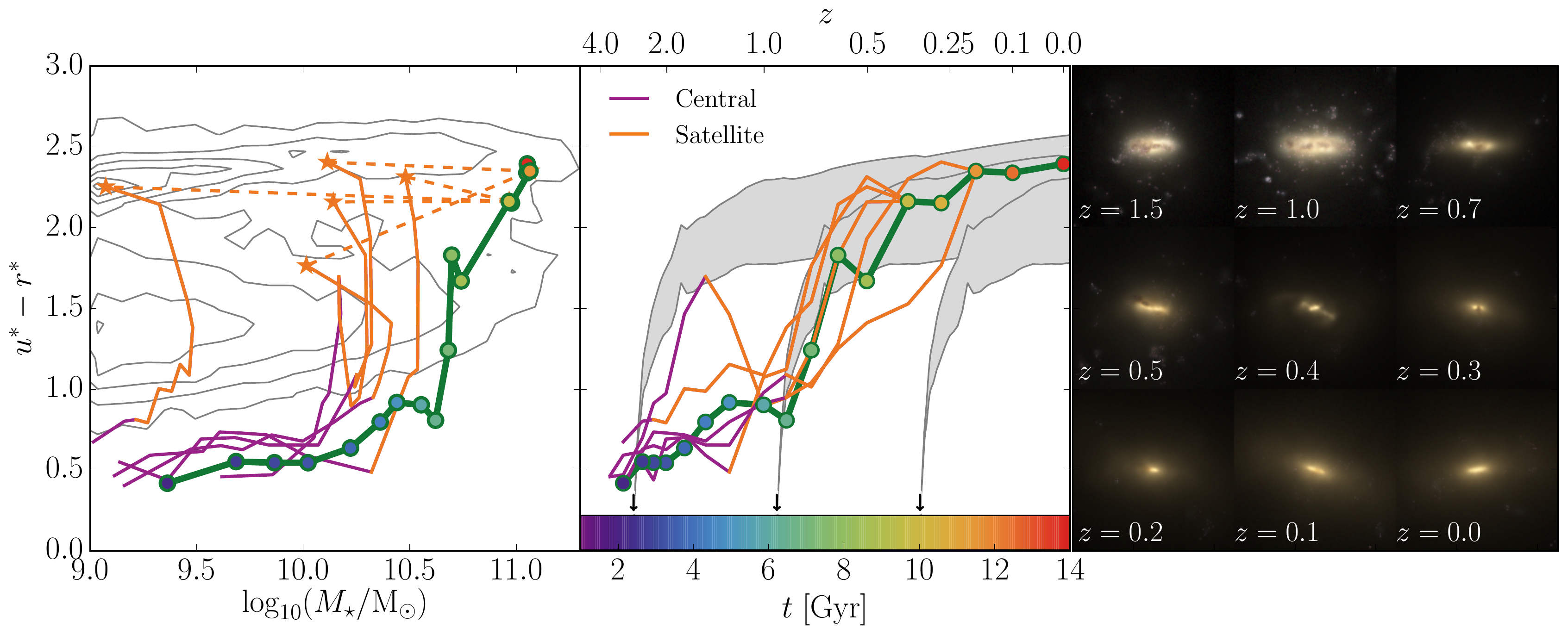}  

 \caption{Same as Fig.~\ref{fig:lowmass_track}, but for a massive galaxy (green track \& filled circles) and galaxies that merge with it (thin tracks). Tracks for merging galaxies (merging with $M_\star > 10^{9} \; {\rm M_\odot}$) are coloured \textit{purple} when they are centrals, and \textit{orange} when they are satellites as indicated in the legend; a star identifies the last snapshot before the satellite merges with the massive galaxy, and the track is linked to that of the massive galaxy at the following snapshot by a dashed line. The right panel shows edge-on $gri$-composite images of the central galaxy of side length $40$ pkpc, at various redshifts labelled in each separate panel.}
\label{fig:massive_track}
 \end{figure*}
 
\begin{figure}
\includegraphics[width=0.46\textwidth]{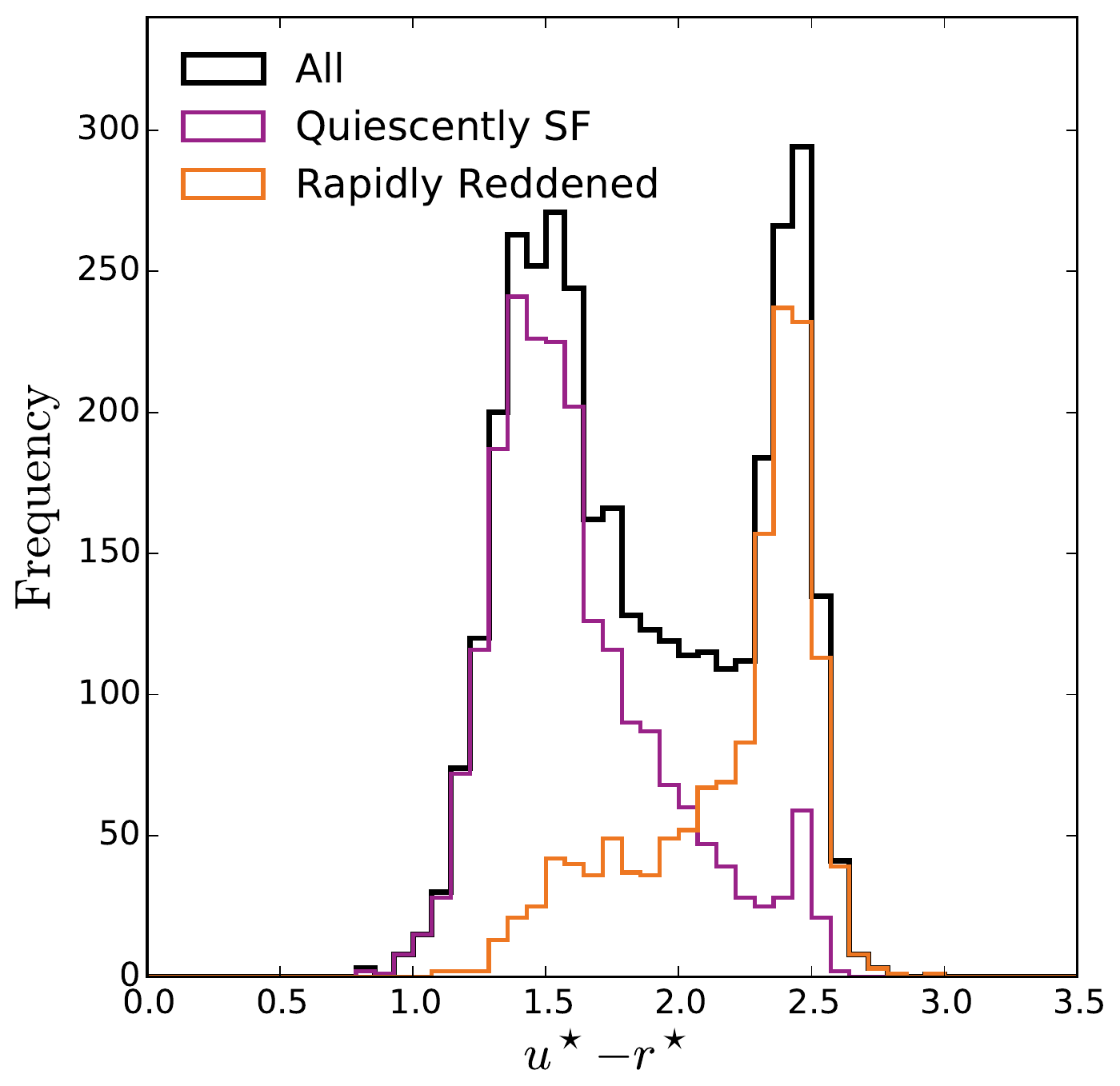}  

 \caption{\us{}-\rs{} colour distribution for $M_\star > 10^{10}{\rm M_\odot}$ galaxies at $z=0$. Galaxies identified as \lq quiescently star-forming \rq are plotted in \textit{purple}, while those that underwent a rapid colour transition to the red are plotted in \textit{orange}. The combined distribution is plotted in black. We see that the quiescently star-forming galaxies predominately inhabit the present-day blue cloud, but with a tail to red colours. Galaxies that underwent a rapid reddening ($\Delta(\hbox{\us-\rs})> 0.8$ in 2 Gyr) are predominately red at $z=0$, but the distribution has a blue tail resulting from recent star formation.}
 \label{fig:class}
 
\end{figure}

\begin{figure}
 \includegraphics[width=0.99\columnwidth]{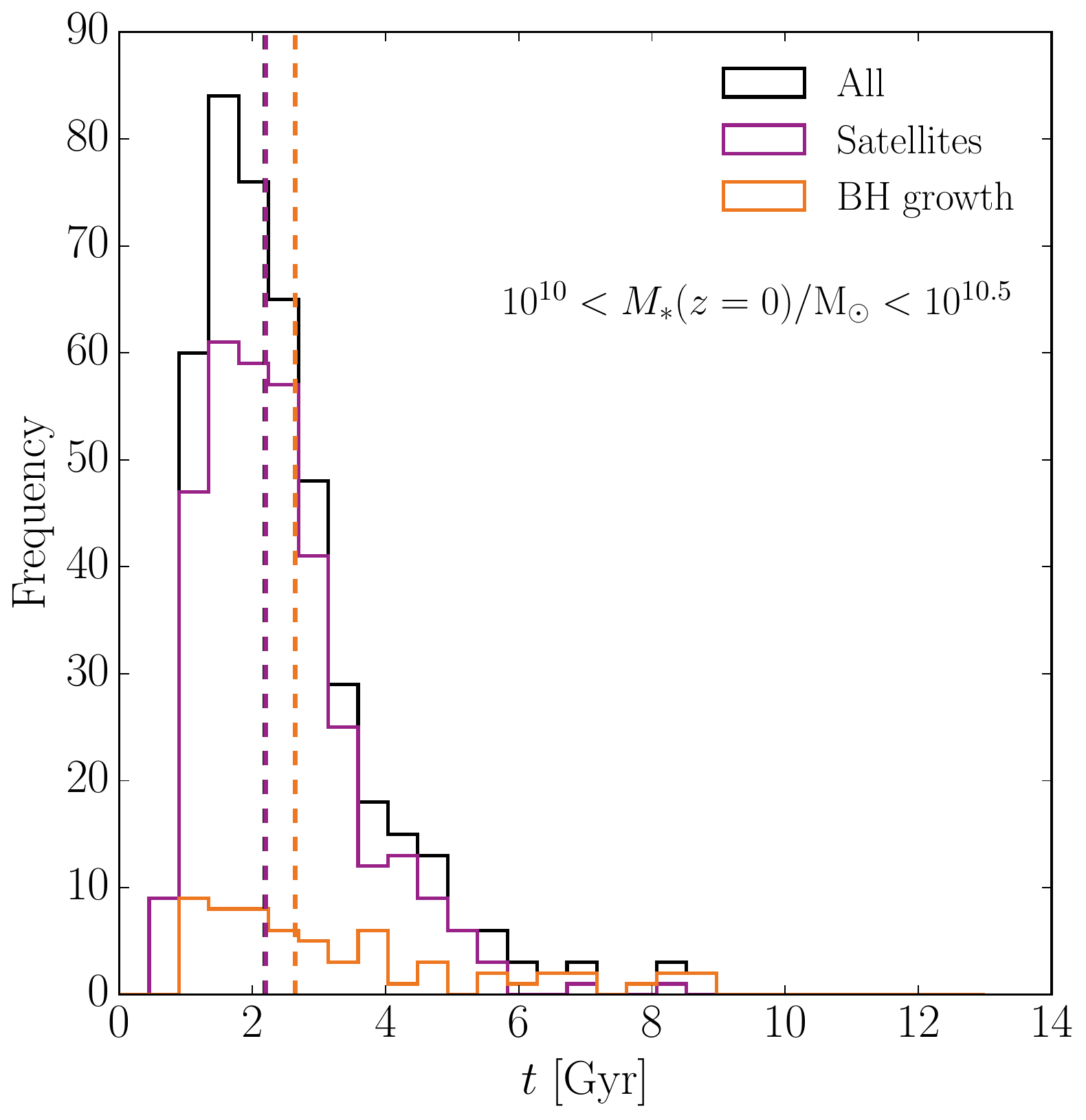}
 \includegraphics[width=0.99\columnwidth]{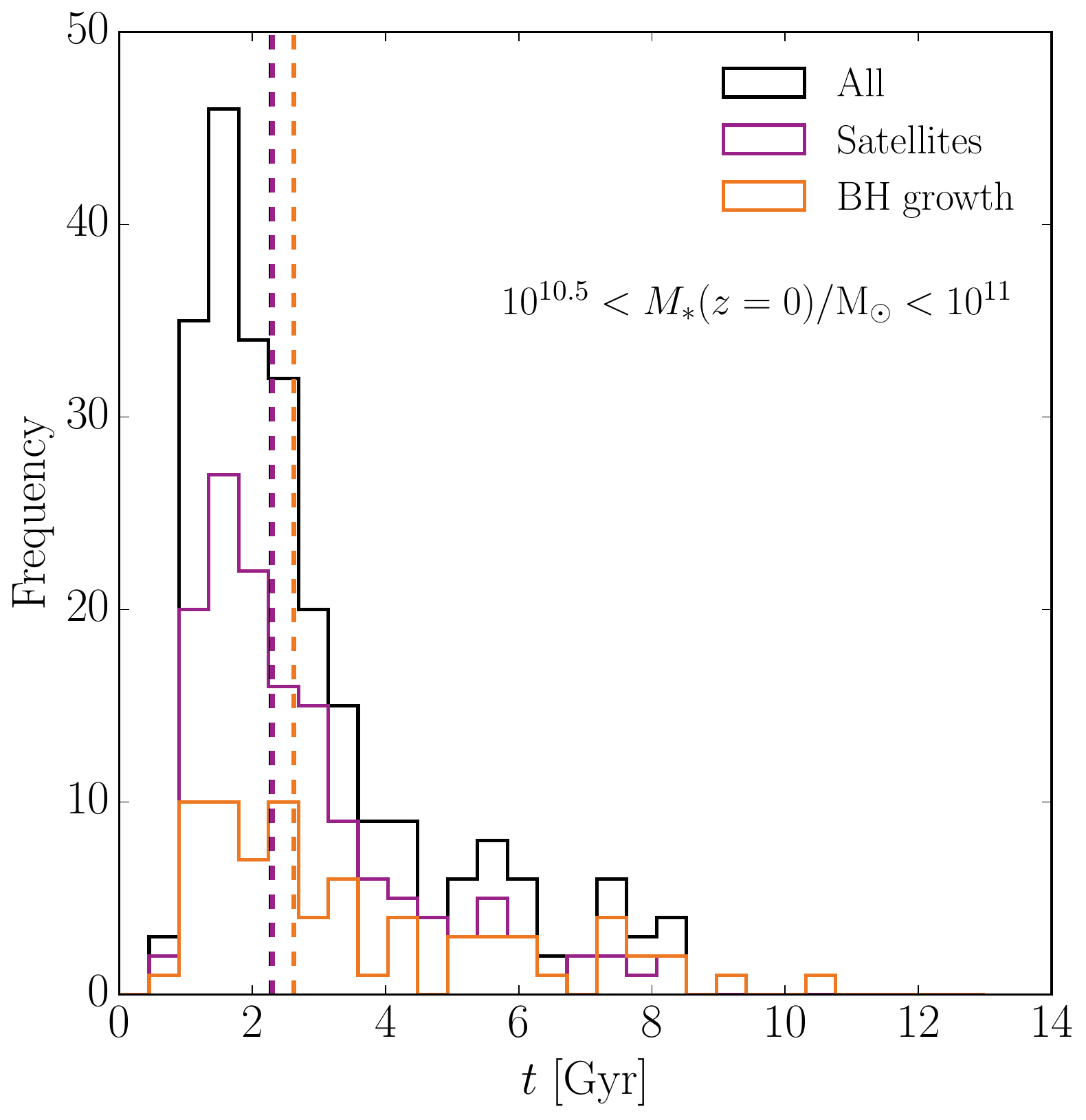}
 \caption{Histograms of the time interval galaxies spent crossing the green valley from being blue to becoming red, $\Delta t_{\rm green}$ (see text). The black histogram is for all galaxies in the current mass selection, {\em purple} histogram for galaxies that are satellites at $z=0$, \textit{orange} histogram for galaxies whose black holes grew by more than a factor of 1.5 while crossing the green valley. Median values for the selections are plotted as dashed lines with the corresponding colour. {\em Top panel} shows galaxies selected in the $z=0$ mass range $10^{10} < M_\star/{\rm M_\odot} < 10^{10.5}$, with the {\em bottom panel} corresponding to $10^{10.5} < M_\star/{\rm M_\odot} < 10^{11}$. Satellite galaxies dominate in both mass ranges. The transition time-scale for a blue galaxy to turn red is typically $~\lesssim 2$~Gyr, which corresponds to the time a blue population of stars reddens passively, as seen in Fig.\ref{fig:lowmass_track}.}
 \label{fig:speed2}
 \end{figure}

\begin{table}
\begin{center}
\caption{Properties of the galaxies plotted as main tracks in Figs.~\ref{fig:lowmass_track}-\ref{fig:massive_track}. The Symbol/Figure is given to identify the galaxies on the figures. For each galaxy we quote the unique galaxy identifier (GalaxyID) taken from the \eagle{} public database \citep{McAlpine15}, the $z=0$ black hole mass ($M_{\bullet}$), and indicate whether a galaxy was ever classified as a satellite (y) or not (n).}
\label{tab:props}
\begin{tabular}{lccc}
\hline
Sym./Fig. & GalaxyID & $M_{\bullet}/{\rm M_\odot}$ & Satellite \\
\hline
Circle/\ref{fig:lowmass_track} & 18169630 & $7.09\times10^{6}$ & n \\ 
Square/\ref{fig:lowmass_track} & 15829793 & $1.03\times10^{8}$ & n\\
Triangle/\ref{fig:lowmass_track} & 14096270 & $6.00\times10^{7}$ & n\\
Circle/\ref{fig:massive_track} &  15197399 & $1.84\times10^{8}$ & y\\
\hline
\end{tabular}
\end{center}
\end{table}

We have examined a large number of tracks of individual galaxies in (\us-\rs, $M_\star$) space and have identified three generic tracks of central galaxies that we illustrate in Fig.~\ref{fig:lowmass_track}. In Fig.~\ref{fig:massive_track} we also show the track of a central galaxy that is very massive at $z=0$ ($M_\star \sim 10^{11}{\rm M_\odot}$), to illustrate individual tracks of satellites that merge with it. More details of the four galaxies tracked in these panels are given in Table~\ref{tab:props}. The time-scale over which galaxies transition to the red sequence is compared to that of a passively evolving population, by plotting (single) star-burst tracks initiated at different times (grey curves in middle panels).

The blue track in Fig.~\ref{fig:lowmass_track} is for a galaxy that remains in the blue cloud down to $z=0$, forming stars in a blue disc that grows in time. As its sSFR decreases with time, and the contribution of an older population of stars becomes more important, it slowly reddens, with $\Delta(\hbox{\us-\rs})\approx 0.2$ from $z=1$ to $z=0$ (elapsed time $\approx 8$~Gyr). We see from Table~\ref{tab:props} and Fig.~\ref{fig:bh} that the black hole mass is $\approx 0.6$~dex lower than the median value for its stellar mass (at approximately the $3$rd percentile of galaxies for that $M_\star$), suggesting low levels of black hole feedback in the galaxy's history. 

The red track in Fig.~\ref{fig:lowmass_track} corresponds to a galaxy that becomes red more rapidly, reddening by $\Delta(\hbox{\us-\rs})\approx 1$ in $\approx 2$~Gyr, joining the red sequence at $z=2$. From then on, its colour or stellar mass hardly changes; it has been a very compact elliptical galaxy since at least $z=1$, maintaining a stellar half-mass radius of $< 2.3$ pkpc. The similar rate of colour transition for this galaxy to that of an instantaneous starburst suggests rapid quenching of star formation.  Considering the black hole mass in Table~\ref{tab:props} and the bottom-left panel Fig.~\ref{fig:bh} shows that the galaxy has a central black hole mass $\approx 0.6$~dex higher than the median black-hole stellar mass relation (at approximately the $99$th percentile of galaxies for that $M_\star$), suggestive of black hole quenching. 

The thick green track shows a galaxy that reddens at a similar rate
($\Delta(\hbox{\us-\rs})\approx 1.5$ in $\approx 2$~Gyr) at $z\approx
1$, after undergoing a near-equal mass merger (thin green line
indicates the track of the other galaxy) leading to significant AGN growth. 
It then changes morphology from being a disturbed disc to a compact
elliptical at $z=0.5$. At redshift $z=0.25$, it starts forming stars
again, turning blue ($\Delta(\hbox{\us-\rs})\approx -1$) over $\approx
2$~Gyr. By $z=0$,it has grown significantly in size, with a prominent
bulge and an extended distribution of stars around it. The value of
$M_{\bullet}$ in Table~\ref{tab:props} puts the black hole mass at
approximately the 75th percentile of galaxies for that stellar mass. 

Note that the black hole masses for galaxies selected in Fig.~\ref{fig:lowmass_track} (and listed in Table~\ref{tab:props}) are all in the upper or lower quartiles for their stellar mass. While galaxies with average black hole masses have diverse histories, more extreme $M_{\bullet}$ values can be indicative of the certain types of track represented here. The quiescently star forming and quenched galaxies are both associated with particularly low and high levels of black hole growth respectively. The rejuvenated galaxy has a high black hole mass, though less extreme. The black hole growth is associated with a quenching event at $z\approx 1$, but the galaxy re-accretes gas and is able to recommence prolonged ($>2$ Gyr) star formation in spite of its high black hole mass. We saw in Fig. \ref{fig:bh} that for $z=0$ and at these masses the overall correlation between galaxy colour and the residuals of the $M_{\bullet}$-$M_\star$ relation is rather weak.   


The massive $z=0$ galaxy in Fig.~\ref{fig:massive_track} is a blue star-forming disc until just below $z=1$, after which it becomes red and evolves into an elongated elliptical. Thin lines show the tracks of five galaxies that merge with it, with the line colour changing from purple to orange while these galaxies become satellites. Examination of these tracks reveals that while some galaxies become red when they are still centrals, most galaxies quench after being identified as satellites. This suggests that satellite identification is a good predictor of colour change in \eagle{}, particularly for galaxies falling into a more massive halo. The satellite tracks exhibit a characteristic shape of rapid quenching followed by stellar mass loss, as the galaxy is stripped and eventually merges. We also see that the central galaxy exhibits quite a stochastic colour evolution compared to those in Fig.~\ref{fig:lowmass_track}. This is perhaps due to the higher frequency of satellite interactions and mergers for the higher mass halo represented in Fig.~\ref{fig:massive_track}, with only the rejuvenated track of Fig.~\ref{fig:lowmass_track} showing (two) mergers with satellites of $M_\star > 10^{9}{\rm M_\odot}$.

The heterogeneous colour evolution of galaxies illustrated in Fig.~\ref{fig:lowmass_track} and \ref{fig:massive_track} are consistent with observations that the green valley population is diverse \citep[e.g.][]{Cortese09, Schawinski14}. In particular, \citet{Cortese09} also find examples of galaxies moving off the red sequence through re-accretion of gas reminiscent of the green track in Fig.~\ref{fig:lowmass_track}.

To quantify the fraction of galaxies that undergo a rapid colour transformation, we trace the main progenitors of all galaxies with mass $M_\star > 10^{10}{\rm M_\odot}$ at $z=0$ back in time (a sample of $\approx 3000$ galaxies). We find that the fraction of galaxies with at least one $\Delta(\hbox{\us-\rs}) > 0.8$ change over any 2~Gyr period in their history is $\approx$~40~per cent. We will refer to such galaxies as rapidly reddening galaxies; they tend to follow a track similar to the red track in Fig.\ref{fig:lowmass_track}. Of the galaxies that redden quickly, $\approx 1.6$~per cent undergo a colour change to the {\em blue} of $\Delta(\hbox{\us-\rs}) < -0.8$ over a 2~Gyr period. We will refer to this small fraction of galaxies as \lq rejuvenated\rq; they follow a track similar to the green track in Fig.~\ref{fig:lowmass_track}\footnote{Galaxies may also undergo slower colour transitions due to secular evolution, and the number density of galaxies does not remain constant because of mergers. These quoted fractions therefore inevitably depend on how galaxies are selected. {\em i.e.} the colour choice we made in Eq.~\ref{eq:select}.}. Galaxies that do not ever undergo such a rapid reddening event make-up 60~per cent of the sample. We will refer to these as \lq quiescently star-forming\rq\ galaxies; they follow a track similar to the blue track in Fig.\ref{fig:lowmass_track}.

Fig.~\ref{fig:class} shows the distribution of $z=0$ colours for $10^{10} < M_\star/{\rm M_\odot} < 10^{10.5}$ galaxies classified as having undergone a rapid transformation to redder colour (orange histogram), and those that never underwent such a rapid reddening (quiescently star-forming galaxies, purple histogram). A small fraction of galaxies become red without ever experiencing a rapid reddening event: this is the tail of the purple histogram towards red \us-\rs{} colour. Similarly, there is a tail to blue \us-\rs{} colour in the orange histogram, representing galaxies that underwent a rapid colour transition to the red, followed by more recent star formation turning them blue once more. The fraction of these galaxies is much higher than the 1.6~per cent of galaxies we classified as \lq rejuvenated\rq: the majority of galaxies that are blue now but were red in the past ($\approx 10 \%$ of total), became blue more gradually than the $\Delta(\hbox{\us-\rs})=-0.8$  over 2~Gyr that we used to define \lq rejuvenated galaxies\rq. 

Galaxies must become blue rapidly or traverse the entire green valley to meet these criteria. However, it is also interesting to note the probability that a galaxy identified in the green valley is on a bluer trajectory in the colour-$M_\star$ plane. We identify this for green valley galaxies at $z<2$, enforcing that a galaxy must undergo a monotonic colour change of $\Delta(\hbox{\us-\rs})<-0.05$ to be deemed significant \citep[above the level of photometric error, e.g.][]{Padmanabhan08}. We find that the \eagle{} green valley galaxies have a $17\%$ chance of being on a significantly blue trajectory. Conversely, we find the probability of a green galaxy being on a significantly red trajectory to be $75\%$, taking $\Delta(\hbox{\us-\rs})>0.05$ as the criteria for being significant.

The tracks of individual galaxies enable us to characterise the colour transition time-scale of a galaxy by the time interval, $\Delta t_{\rm green}$, it spent in the green valley on its way from the blue cloud to the red sequence. We calculate $\Delta t_{\rm green}$ as follows: using Eq.~(\ref{eq:select}) we select red galaxies at $z=0$ and trace their main progenitors back in time to identify the last time ($t_1$) they became red and the last time ($t_2$) they were blue.  Histograms of colour transition times, $\Delta t_{\rm green}\equiv t_1-t_2$, for galaxies that at $z=0$ have $10^{10} < M_\star/{\rm M_\odot} < 10^{10.5}$ and $10^{10.5} < M_\star/{\rm M_\odot} < 10^{11}$ are plotted in Fig.~\ref{fig:speed2} (top and bottom panels, respectively).

The mode of the colour transition time distribution is $\approx 1.5$~Gyr, with a median of $\approx 2$~Gyr, mostly independent of whether quenching is likely due to becoming a satellite or AGN activity (purple and  orange histograms, respectively). This is the time-scale for a passively evolving blue population of stars to turn red, as can be seen from Fig.\ref{fig:lowmass_track}. Strikingly, there is a very long tail to high values in the distribution of $\Delta t_{\rm green}$ as inefficient quenching allows a small fraction of galaxies to spend a long time in the green valley before eventually turning red, whether due to becoming a satellite or hosting an AGN. Though quenched galaxies are more prevalent in high-mass halos, the colour transition time-scales show little dependence on halo mass. Despite this, the longest time-scales we measure are for halo masses $< 10^{13}$ M$_\odot$.

 The time-scales for colour transition and the quenching of star formation are clearly linked, however colour transition times are longer due to the passive evolution of stellar populations. This is illustrated by the grey curves in Fig. \ref{fig:lowmass_track} and \ref{fig:massive_track}, showing that the colour transition time for an SSP is $\approx 2$ Gyr. The quenching time-scales of observed satellites presented by e.g. \citet{Muzzin14} and \citet{Wetzel13} are significantly shorter, typically $\lesssim 0.5$ Gyr and $\lesssim 0.8$ Gyr respectively. Although the \us-\rs{} colour index alone is not sensitive enough to resolve these quenching times, the typical colour transition times of \eagle{} galaxies are consistent with such rapid quenching. 

However, there are cluster studies in the $0 < z < 0.5$ redshift range
that infer longer timescales  \citep[$\gtrapprox 1$ Gyr,
e.g.][]{VonDerLinden10, Vulcani10, Haines13}. Fig.~\ref{fig:speed2}
does show a tail to longer quenching times for both AGN and
satellites, though they are not typical at low redshift. Ultimately,
the different observational tracers and modelling used to infer
quenching times are still subject to significant systematics that may
explain the different measurements \citep{McGee14}. In particular,
colour transition and quenching timescales are not equivalent, and
galaxies may move between the red and blue populations when observed
in optical and UV colour \citep[e.g.]{Cortese12}. An analysis of the
physical quenching timescale and its evolution is left to a future
study.

%% file: Summary.tex
\section{Conclusions}
\label{sec:summary}
We have investigated the evolution and origin of the colours of galaxies in the \eagle\ cosmological hydrodynamical simulation \citep{Schaye15, Crain15}. We apply the single population synthesis models from \cite{bc03} to model galaxy colours in the absence of dust, as described by \citet{Trayford15}. We also use galaxy merger trees to trace descendants as well as main progenitors through time.

The \us-\rs\ vs $M_\star$ diagram is bimodal at redshift $z=0$, with a clearly defined red sequence of quenched galaxies, and a blue cloud of star-forming galaxies. The scatter and slope of the red sequence are both determined mainly by stellar metallicity, while the normalisation additionally depends on stellar age (Fig. \ref{fig:col}). The scatter in the blue cloud, in contrast, is mostly due to scatter in the specific star formation rate at fixed stellar mass (${\rm sSFR}=\dot M_\star/M_\star$). The slope of blue cloud colours versus $M_\star$ is similar to that of the red sequence, but as their origins are different, this is coincidental. At higher $z$, both colour sequences become bluer, and the red sequence becomes less populated until it has mostly disappeared by $z=2$.

From studying the evolution of \eagle{} galaxies in \us{}-\rs{} and $M_\star$, we note that in general:

\begin{itemize}
\item Galaxies in \eagle\ turn red either because they become satellites (mainly at lower masses, see Fig. \ref{fig:sat}) or because of feedback from their central supermassive black hole (mainly for more massive galaxies, see Fig. \ref{fig:bh}). As a consequence, the red sequence builds-up from both the low-mass and high-mass sides simultaneously, with the low-mass red galaxies being satellites of the massive red centrals that are quenched by their AGN. This results in a dearth of red galaxies at intermediate mass, $M_\star\sim 10^{10}$~M$_\odot$, at $z\approx 1$ - such galaxies are too low mass to host a massive black hole, but too massive for a large fraction of them to be satellites in the \eagle{} volume. While we believe the existence of such a deficit is unlikely to change with increased simulation volume, it should be noted that the limited volume and lack of large scale power in the \eagle{} 100$^3$ Mpc$^3$ simulation may affect the depth of the deficit.
\item The colour evolution in the blue cloud is driven by the decrease in the sSFR rates of star-forming galaxies with cosmic time.
\item The characteristic time scale for galaxies to cross the green valley, from the blue cloud to the red sequence (Fig. \ref{fig:speed2}), is $\Delta t_{\rm green}\approx$~2~Gyr, mostly independent of galaxy mass and cause of the quenching. It is determined by the rate of evolution of a passive population of blue stars to the red. This timescale is consistent with rapid or instantaneous quenching of star formation, as inferred from observations of satellite galaxies by \citet{Muzzin14}. The distribution of $\Delta t_{\rm green}$ has an extended tail to $\sim 10$~Gyr: a small fraction of galaxies remain green for a long time. However, most galaxies spend only a short time, $\Delta t_{\rm green}\lesssim 2$~Gyr, in the green valley - it is not easy being green.
\end{itemize}

We identified three characteristic tracks that galaxies follow in the \us-\rs vs $M_\star$ diagram (Figs. \ref{fig:lowmass_track} and \ref{fig:massive_track}). {\em Quiescently star-forming galaxies} remain in the blue cloud at all times, without sudden reddening episodes of $\Delta(\hbox{\us-\rs}) > 0.8$ in any 2~Gyr interval. Nearly 60 per cent of galaxies with stellar mass at $z=0$ greater than $M_\star = 10^{10}$~${\rm M_\odot}$ fall into this category (see Fig. \ref{fig:class}). The remaining 40~per cent of galaxies do undergo such sudden episodes of star formation suppression. The majority of these {\em rapidly reddened galaxies} move onto the red sequence permanently as per the evolutionary picture of \citet{Faber07}, however we find that 1.6~per cent undergo an episode in which star formation causes the galaxy to change colour to the blue again, having $\Delta(\hbox{\us-\rs}) < -0.8$ over a 2~Gyr period (e.g. Fig \ref{fig:lowmass_track}). The fraction of such {\em rejuvenated galaxies} is thus very small. Nevertheless, a much larger fraction of the galaxies that at $z=0$ are blue were red in the past: the rate of colour transition of galaxies to the blue is generally significantly slower than the quenching timescale. We also find that the fraction of green valley galaxies on blue trajectories (where $\Delta(\hbox{\us-\rs})<-0.05$) at a given instance from $z<2$ is larger still at $17\%$, implying that only a subset transition completely from red to blue and remain there.